%% file: main.tex
\documentclass[sigplan,screen]{acmart} 

\AtBeginDocument{%
  \providecommand\BibTeX{{%
    \normalfont B\kern-0.5em{\scshape i\kern-0.25em b}\kern-0.8em\TeX}}}

\setcopyright{acmlicensed}
\acmPrice{15.00}
\acmDOI{10.1145/3437992.3439922}
\acmYear{2021}
\copyrightyear{2021}
\acmSubmissionID{poplws21cppmain-p21-p}
\acmISBN{978-1-4503-8299-1/21/01}
\acmConference[CPP '21]{Proceedings of the 10th ACM SIGPLAN International Conference on Certified Programs and Proofs}{January 18--19, 2021}{Virtual, Denmark}
\acmBooktitle{Proceedings of the 10th ACM SIGPLAN International Conference on Certified Programs and Proofs (CPP '21), January 18--19, 2021, Virtual, Denmark}

\input{includes}

\input{macros}

\begin{document}

\title{Formalizing Category Theory in Agda}

\author{Jason Z. S. Hu}
\email{zhong.s.hu@mail.mcgill.ca}
\affiliation{%
  \institution{McGill University}
  \streetaddress{McConnell Engineering Bldg. 3480 University St.,}
  \city{Montr\'eal}
  \state{Qu\'ebec}
  \country{Canada}
  \postcode{H3A 0E9}
}

\author{Jacques Carette}
\email{carette@mcmaster.ca}
\affiliation{%
  \institution{McMaster University}
  \streetaddress{Room ITB-168,  1280 Main Street West}
  \city{Hamilton}
  \state{Ontario}
  \country{Canada}
}


\begin{abstract}
The generality and pervasiveness of category theory in modern mathematics makes it
a frequent and useful target of formalization.  It is however quite challenging
to formalize, for a variety of reasons.  Agda currently (i.e. in 2020) does not have
a standard, working formalization of category theory.  We document our work on
solving this dilemma.  The formalization revealed a number of potential design
choices, and we present, motivate and explain the ones we picked.  In
particular, we find that alternative definitions or alternative proofs from
those found in standard textbooks can be advantageous, as well as ``fit''
Agda's type theory more smoothly. Some definitions regarded as equivalent in
standard textbooks turn out to make different ``universe level'' assumptions,
with some being more polymorphic than others. We also pay close attention to
engineering issues so that the library integrates well with Agda's own standard
library, as well as being compatible with as many of supported type theories
in Agda as possible.
\end{abstract}




\maketitle

\section{Introduction}

There have been many formalizations of category
theory~\cite{Awodey:2010:CT:2060081,maclane:71} in many different proof assistants,
over more than $25$ years~\cite[etc.]{timany_et_al:LIPIcs:2016:6000,huet2000constructive,UniMath,The_mathlib_Community_2020,ahrens_kapulkin_shulman_2015,10.1007/978-3-319-08970-6_18,copumpkin,coq:category,Category3-AFP,MonoidalCategory-AFP,Bicategory-AFP}.
All of them embody \emph{choices}; some were forced by the ambient logic of the
host system, others were pragmatic decisions, some were philosophical stances, 
while finally others were simply design decisions.

Category theory is often picked as a challenge, as it is
both be quite amenable to formalization and to involve many non-trivial
decisions that can have drastic effects on the usability and effectiveness of
the results~\cite{10.1007/978-3-319-08970-6_18}.  With the
rapid rise in the use of category theory
as a tool in computer science, and with the advent of \emph{applied} category
theory, having a stable formalization in the \emph{standard library} of one's
favourite proof assistant becomes necessary.

Our journey started as the authors were trying to keep the ``old'' category
theory library for Agda~\cite{copumpkin} alive.  Unfortunately, as Agda~\cite{agda} evolved,
some of the features used in that library were no longer well-supported, and
eventually the library simply stopped working.  As it became clear that simply
continuing to patch that library was no longer viable, a new version was in
order.

This gave us the opportunity to revisit various design decisions of the earlier
implementation --- which we will document.  We also wanted to preserve as much
formalization effort as possible, while also use
language features introduced in Agda 2.6+ like generalized variables  and revise the theoretical foundation which
the library relies on. This new version is then partly a
``port'' of the previous one to current versions of Agda, but also heavily
refactored, including some large changes in design.

Our principal theoretical contribution is to show that \emph{setoid-based
proof-relevant category theory} works just as well as various other
``flavours'' of category theory by supporting a large number of definitions and
theorems.  Our main engineering contribution is a coherent set of design
decisions for a widely reusable and working library of category theory in Agda,
freely available%
\footnote{at \url{https://github.com/agda/agda-categories}}.

This paper is structured as follows. In~\Cref{sec:basics}, we discuss our
global design choices. We discuss the rationale behind non-strictness,
proof-relevance, hom-setoids, universe polymorphism, (not) requiring extra laws and
concepts as record types. In~\Cref{sec:pr-def}, we give examples on how
proof-relevance drives us to find concepts in an alternative
way. In~\Cref{sec:discuss}, we discuss other design
decisions and some efficiency issues.
In~\Cref{sec:related}, we compare category theory libraries
in other systems. Finally, we conclude in~\Cref{sec:concl}.

For reasons of space, we have to make some assumptions of our
readership, namely that they are familiar with:
\begin{enumerate}
\item category theory,
\item dependent type theory,
\item formalization, and
\item proof assistants (e.g. familiarity with Agda and a 
passing knowledge of other systems).
\end{enumerate}

\input{formal}

\input{method}

\input{survey}

\input{discussion}

\input{related}

\section{Conclusion and Future Work}\label{sec:concl}

We implemented proof-relevant category theory in Agda, successfully. The concepts
covered, and the theorems proved, are quite broad.  We did not
find any real barrier to doing so --- strictness and hom-sets are not necessary
features of modern category theory.  We did find that some definitions work better
than others, which we have explained in detail.  Comparing with other libraries,
we find that ours covers quite similar grounds, and often more.

We are still actively developping this library --- many theorems of classical
category theory remain; both bicategory theory and enriched category
theory are being built up. Some work has been done on ``negative thinking''
($-2$-categories, etc) and should be extended.  Both double categories and higher
categories are still awaiting, along with multicategories, PROPs, operads and
polycategories. We also intend to move parts of this library to the standard
library.

Performance needs another look. Even after some optimizations were performed,
it still takes more memory and time to typecheck than we would prefer. Having said
that, development can easily be done on a normal laptop, so the problem is
not severe, unlike with other libraries.

\begin{acks}
  We would like to thank Sandro Stucki, Reed Mullanix, Nathan van Doorn, and many
  others for discussions and contributing to the library. We are also grateful to the anonymous reviewers
  for their inspirational suggestions.

  This work was supported by the National Sciences and Engineering Research Council of Canada. 
\end{acks}



\end{document}

%% file: includes.tex
\usepackage[T1]{fontenc} 
\usepackage[utf8]{inputenc}

\usepackage{amsmath}
\usepackage[draft]{minted}
\usepackage{tikz-cd}
\usepackage{array}
\usepackage{amssymb}
\usepackage{pifont}
\usepackage{cleveref}
\usepackage[htt]{hyphenat}
\usepackage{xspace}
\usepackage{microtype}

\makeatletter
\def\PYGdefault@reset{\let\PYGdefault@it=\relax \let\PYGdefault@bf=\relax%
    \let\PYGdefault@ul=\relax \let\PYGdefault@tc=\relax%
    \let\PYGdefault@bc=\relax \let\PYGdefault@ff=\relax}
\def\PYGdefault@tok#1{\csname PYGdefault@tok@#1\endcsname}
\def\PYGdefault@toks#1+{\ifx\relax#1\empty\else%
    \PYGdefault@tok{#1}\expandafter\PYGdefault@toks\fi}
\def\PYGdefault@do#1{\PYGdefault@bc{\PYGdefault@tc{\PYGdefault@ul{%
    \PYGdefault@it{\PYGdefault@bf{\PYGdefault@ff{#1}}}}}}}
\def\PYGdefault#1#2{\PYGdefault@reset\PYGdefault@toks#1+\relax+\PYGdefault@do{#2}}

\expandafter\def\csname PYGdefault@tok@gd\endcsname{\def\PYGdefault@tc##1{\textcolor[rgb]{0.63,0.00,0.00}{##1}}}
\expandafter\def\csname PYGdefault@tok@gu\endcsname{\let\PYGdefault@bf=\textbf\def\PYGdefault@tc##1{\textcolor[rgb]{0.50,0.00,0.50}{##1}}}
\expandafter\def\csname PYGdefault@tok@gt\endcsname{\def\PYGdefault@tc##1{\textcolor[rgb]{0.00,0.27,0.87}{##1}}}
\expandafter\def\csname PYGdefault@tok@gs\endcsname{\let\PYGdefault@bf=\textbf}
\expandafter\def\csname PYGdefault@tok@gr\endcsname{\def\PYGdefault@tc##1{\textcolor[rgb]{1.00,0.00,0.00}{##1}}}
\expandafter\def\csname PYGdefault@tok@cm\endcsname{\let\PYGdefault@it=\textit\def\PYGdefault@tc##1{\textcolor[rgb]{0.25,0.50,0.50}{##1}}}
\expandafter\def\csname PYGdefault@tok@vg\endcsname{\def\PYGdefault@tc##1{\textcolor[rgb]{0.10,0.09,0.49}{##1}}}
\expandafter\def\csname PYGdefault@tok@m\endcsname{\def\PYGdefault@tc##1{\textcolor[rgb]{0.40,0.40,0.40}{##1}}}
\expandafter\def\csname PYGdefault@tok@mh\endcsname{\def\PYGdefault@tc##1{\textcolor[rgb]{0.40,0.40,0.40}{##1}}}
\expandafter\def\csname PYGdefault@tok@go\endcsname{\def\PYGdefault@tc##1{\textcolor[rgb]{0.53,0.53,0.53}{##1}}}
\expandafter\def\csname PYGdefault@tok@ge\endcsname{\let\PYGdefault@it=\textit}
\expandafter\def\csname PYGdefault@tok@vc\endcsname{\def\PYGdefault@tc##1{\textcolor[rgb]{0.10,0.09,0.49}{##1}}}
\expandafter\def\csname PYGdefault@tok@il\endcsname{\def\PYGdefault@tc##1{\textcolor[rgb]{0.40,0.40,0.40}{##1}}}
\expandafter\def\csname PYGdefault@tok@cs\endcsname{\let\PYGdefault@it=\textit\def\PYGdefault@tc##1{\textcolor[rgb]{0.25,0.50,0.50}{##1}}}
\expandafter\def\csname PYGdefault@tok@cp\endcsname{\def\PYGdefault@tc##1{\textcolor[rgb]{0.74,0.48,0.00}{##1}}}
\expandafter\def\csname PYGdefault@tok@gi\endcsname{\def\PYGdefault@tc##1{\textcolor[rgb]{0.00,0.63,0.00}{##1}}}
\expandafter\def\csname PYGdefault@tok@gh\endcsname{\let\PYGdefault@bf=\textbf\def\PYGdefault@tc##1{\textcolor[rgb]{0.00,0.00,0.50}{##1}}}
\expandafter\def\csname PYGdefault@tok@ni\endcsname{\let\PYGdefault@bf=\textbf\def\PYGdefault@tc##1{\textcolor[rgb]{0.60,0.60,0.60}{##1}}}
\expandafter\def\csname PYGdefault@tok@nl\endcsname{\def\PYGdefault@tc##1{\textcolor[rgb]{0.63,0.63,0.00}{##1}}}
\expandafter\def\csname PYGdefault@tok@nn\endcsname{\let\PYGdefault@bf=\textbf\def\PYGdefault@tc##1{\textcolor[rgb]{0.00,0.00,1.00}{##1}}}
\expandafter\def\csname PYGdefault@tok@no\endcsname{\def\PYGdefault@tc##1{\textcolor[rgb]{0.53,0.00,0.00}{##1}}}
\expandafter\def\csname PYGdefault@tok@na\endcsname{\def\PYGdefault@tc##1{\textcolor[rgb]{0.49,0.56,0.16}{##1}}}
\expandafter\def\csname PYGdefault@tok@nb\endcsname{\def\PYGdefault@tc##1{\textcolor[rgb]{0.00,0.50,0.00}{##1}}}
\expandafter\def\csname PYGdefault@tok@nc\endcsname{\let\PYGdefault@bf=\textbf\def\PYGdefault@tc##1{\textcolor[rgb]{0.00,0.00,1.00}{##1}}}
\expandafter\def\csname PYGdefault@tok@nd\endcsname{\def\PYGdefault@tc##1{\textcolor[rgb]{0.67,0.13,1.00}{##1}}}
\expandafter\def\csname PYGdefault@tok@ne\endcsname{\let\PYGdefault@bf=\textbf\def\PYGdefault@tc##1{\textcolor[rgb]{0.82,0.25,0.23}{##1}}}
\expandafter\def\csname PYGdefault@tok@nf\endcsname{\def\PYGdefault@tc##1{\textcolor[rgb]{0.00,0.00,1.00}{##1}}}
\expandafter\def\csname PYGdefault@tok@si\endcsname{\let\PYGdefault@bf=\textbf\def\PYGdefault@tc##1{\textcolor[rgb]{0.73,0.40,0.53}{##1}}}
\expandafter\def\csname PYGdefault@tok@s2\endcsname{\def\PYGdefault@tc##1{\textcolor[rgb]{0.73,0.13,0.13}{##1}}}
\expandafter\def\csname PYGdefault@tok@vi\endcsname{\def\PYGdefault@tc##1{\textcolor[rgb]{0.10,0.09,0.49}{##1}}}
\expandafter\def\csname PYGdefault@tok@nt\endcsname{\let\PYGdefault@bf=\textbf\def\PYGdefault@tc##1{\textcolor[rgb]{0.00,0.50,0.00}{##1}}}
\expandafter\def\csname PYGdefault@tok@nv\endcsname{\def\PYGdefault@tc##1{\textcolor[rgb]{0.10,0.09,0.49}{##1}}}
\expandafter\def\csname PYGdefault@tok@s1\endcsname{\def\PYGdefault@tc##1{\textcolor[rgb]{0.73,0.13,0.13}{##1}}}
\expandafter\def\csname PYGdefault@tok@sh\endcsname{\def\PYGdefault@tc##1{\textcolor[rgb]{0.73,0.13,0.13}{##1}}}
\expandafter\def\csname PYGdefault@tok@sc\endcsname{\def\PYGdefault@tc##1{\textcolor[rgb]{0.73,0.13,0.13}{##1}}}
\expandafter\def\csname PYGdefault@tok@sx\endcsname{\def\PYGdefault@tc##1{\textcolor[rgb]{0.00,0.50,0.00}{##1}}}
\expandafter\def\csname PYGdefault@tok@bp\endcsname{\def\PYGdefault@tc##1{\textcolor[rgb]{0.00,0.50,0.00}{##1}}}
\expandafter\def\csname PYGdefault@tok@c1\endcsname{\let\PYGdefault@it=\textit\def\PYGdefault@tc##1{\textcolor[rgb]{0.25,0.50,0.50}{##1}}}
\expandafter\def\csname PYGdefault@tok@kc\endcsname{\let\PYGdefault@bf=\textbf\def\PYGdefault@tc##1{\textcolor[rgb]{0.00,0.50,0.00}{##1}}}
\expandafter\def\csname PYGdefault@tok@c\endcsname{\let\PYGdefault@it=\textit\def\PYGdefault@tc##1{\textcolor[rgb]{0.25,0.50,0.50}{##1}}}
\expandafter\def\csname PYGdefault@tok@mf\endcsname{\def\PYGdefault@tc##1{\textcolor[rgb]{0.40,0.40,0.40}{##1}}}
\expandafter\def\csname PYGdefault@tok@err\endcsname{\def\PYGdefault@bc##1{\setlength{\fboxsep}{0pt}\fcolorbox[rgb]{1.00,0.00,0.00}{1,1,1}{\strut ##1}}}
\expandafter\def\csname PYGdefault@tok@kd\endcsname{\let\PYGdefault@bf=\textbf\def\PYGdefault@tc##1{\textcolor[rgb]{0.00,0.50,0.00}{##1}}}
\expandafter\def\csname PYGdefault@tok@ss\endcsname{\def\PYGdefault@tc##1{\textcolor[rgb]{0.10,0.09,0.49}{##1}}}
\expandafter\def\csname PYGdefault@tok@sr\endcsname{\def\PYGdefault@tc##1{\textcolor[rgb]{0.73,0.40,0.53}{##1}}}
\expandafter\def\csname PYGdefault@tok@mo\endcsname{\def\PYGdefault@tc##1{\textcolor[rgb]{0.40,0.40,0.40}{##1}}}
\expandafter\def\csname PYGdefault@tok@kn\endcsname{\let\PYGdefault@bf=\textbf\def\PYGdefault@tc##1{\textcolor[rgb]{0.00,0.50,0.00}{##1}}}
\expandafter\def\csname PYGdefault@tok@mi\endcsname{\def\PYGdefault@tc##1{\textcolor[rgb]{0.40,0.40,0.40}{##1}}}
\expandafter\def\csname PYGdefault@tok@gp\endcsname{\let\PYGdefault@bf=\textbf\def\PYGdefault@tc##1{\textcolor[rgb]{0.00,0.00,0.50}{##1}}}
\expandafter\def\csname PYGdefault@tok@o\endcsname{\def\PYGdefault@tc##1{\textcolor[rgb]{0.40,0.40,0.40}{##1}}}
\expandafter\def\csname PYGdefault@tok@kr\endcsname{\let\PYGdefault@bf=\textbf\def\PYGdefault@tc##1{\textcolor[rgb]{0.00,0.50,0.00}{##1}}}
\expandafter\def\csname PYGdefault@tok@s\endcsname{\def\PYGdefault@tc##1{\textcolor[rgb]{0.73,0.13,0.13}{##1}}}
\expandafter\def\csname PYGdefault@tok@kp\endcsname{\def\PYGdefault@tc##1{\textcolor[rgb]{0.00,0.50,0.00}{##1}}}
\expandafter\def\csname PYGdefault@tok@w\endcsname{\def\PYGdefault@tc##1{\textcolor[rgb]{0.73,0.73,0.73}{##1}}}
\expandafter\def\csname PYGdefault@tok@kt\endcsname{\def\PYGdefault@tc##1{\textcolor[rgb]{0.69,0.00,0.25}{##1}}}
\expandafter\def\csname PYGdefault@tok@ow\endcsname{\let\PYGdefault@bf=\textbf\def\PYGdefault@tc##1{\textcolor[rgb]{0.67,0.13,1.00}{##1}}}
\expandafter\def\csname PYGdefault@tok@sb\endcsname{\def\PYGdefault@tc##1{\textcolor[rgb]{0.73,0.13,0.13}{##1}}}
\expandafter\def\csname PYGdefault@tok@k\endcsname{\let\PYGdefault@bf=\textbf\def\PYGdefault@tc##1{\textcolor[rgb]{0.00,0.50,0.00}{##1}}}
\expandafter\def\csname PYGdefault@tok@se\endcsname{\let\PYGdefault@bf=\textbf\def\PYGdefault@tc##1{\textcolor[rgb]{0.73,0.40,0.13}{##1}}}
\expandafter\def\csname PYGdefault@tok@sd\endcsname{\let\PYGdefault@it=\textit\def\PYGdefault@tc##1{\textcolor[rgb]{0.73,0.13,0.13}{##1}}}

\makeatother

\makeatletter
\def\PYG@reset{\let\PYG@it=\relax \let\PYG@bf=\relax%
    \let\PYG@ul=\relax \let\PYG@tc=\relax%
    \let\PYG@bc=\relax \let\PYG@ff=\relax}
\def\PYG@tok#1{\csname PYG@tok@#1\endcsname}
\def\PYG@toks#1+{\ifx\relax#1\empty\else%
    \PYG@tok{#1}\expandafter\PYG@toks\fi}
\def\PYG@do#1{\PYG@bc{\PYG@tc{\PYG@ul{%
    \PYG@it{\PYG@bf{\PYG@ff{#1}}}}}}}
\def\PYG#1#2{\PYG@reset\PYG@toks#1+\relax+\PYG@do{#2}}

\expandafter\def\csname PYG@tok@gd\endcsname{\def\PYG@tc##1{\textcolor[rgb]{0.63,0.00,0.00}{##1}}}
\expandafter\def\csname PYG@tok@gu\endcsname{\let\PYG@bf=\textbf\def\PYG@tc##1{\textcolor[rgb]{0.50,0.00,0.50}{##1}}}
\expandafter\def\csname PYG@tok@gt\endcsname{\def\PYG@tc##1{\textcolor[rgb]{0.00,0.27,0.87}{##1}}}
\expandafter\def\csname PYG@tok@gs\endcsname{\let\PYG@bf=\textbf}
\expandafter\def\csname PYG@tok@gr\endcsname{\def\PYG@tc##1{\textcolor[rgb]{1.00,0.00,0.00}{##1}}}
\expandafter\def\csname PYG@tok@cm\endcsname{\let\PYG@it=\textit\def\PYG@tc##1{\textcolor[rgb]{0.25,0.50,0.50}{##1}}}
\expandafter\def\csname PYG@tok@vg\endcsname{\def\PYG@tc##1{\textcolor[rgb]{0.10,0.09,0.49}{##1}}}
\expandafter\def\csname PYG@tok@m\endcsname{\def\PYG@tc##1{\textcolor[rgb]{0.40,0.40,0.40}{##1}}}
\expandafter\def\csname PYG@tok@mh\endcsname{\def\PYG@tc##1{\textcolor[rgb]{0.40,0.40,0.40}{##1}}}
\expandafter\def\csname PYG@tok@go\endcsname{\def\PYG@tc##1{\textcolor[rgb]{0.53,0.53,0.53}{##1}}}
\expandafter\def\csname PYG@tok@ge\endcsname{\let\PYG@it=\textit}
\expandafter\def\csname PYG@tok@vc\endcsname{\def\PYG@tc##1{\textcolor[rgb]{0.10,0.09,0.49}{##1}}}
\expandafter\def\csname PYG@tok@il\endcsname{\def\PYG@tc##1{\textcolor[rgb]{0.40,0.40,0.40}{##1}}}
\expandafter\def\csname PYG@tok@cs\endcsname{\let\PYG@it=\textit\def\PYG@tc##1{\textcolor[rgb]{0.25,0.50,0.50}{##1}}}
\expandafter\def\csname PYG@tok@cp\endcsname{\def\PYG@tc##1{\textcolor[rgb]{0.74,0.48,0.00}{##1}}}
\expandafter\def\csname PYG@tok@gi\endcsname{\def\PYG@tc##1{\textcolor[rgb]{0.00,0.63,0.00}{##1}}}
\expandafter\def\csname PYG@tok@gh\endcsname{\let\PYG@bf=\textbf\def\PYG@tc##1{\textcolor[rgb]{0.00,0.00,0.50}{##1}}}
\expandafter\def\csname PYG@tok@ni\endcsname{\let\PYG@bf=\textbf\def\PYG@tc##1{\textcolor[rgb]{0.60,0.60,0.60}{##1}}}
\expandafter\def\csname PYG@tok@nl\endcsname{\def\PYG@tc##1{\textcolor[rgb]{0.63,0.63,0.00}{##1}}}
\expandafter\def\csname PYG@tok@nn\endcsname{\let\PYG@bf=\textbf\def\PYG@tc##1{\textcolor[rgb]{0.00,0.00,1.00}{##1}}}
\expandafter\def\csname PYG@tok@no\endcsname{\def\PYG@tc##1{\textcolor[rgb]{0.53,0.00,0.00}{##1}}}
\expandafter\def\csname PYG@tok@na\endcsname{\def\PYG@tc##1{\textcolor[rgb]{0.49,0.56,0.16}{##1}}}
\expandafter\def\csname PYG@tok@nb\endcsname{\def\PYG@tc##1{\textcolor[rgb]{0.00,0.50,0.00}{##1}}}
\expandafter\def\csname PYG@tok@nc\endcsname{\let\PYG@bf=\textbf\def\PYG@tc##1{\textcolor[rgb]{0.00,0.00,1.00}{##1}}}
\expandafter\def\csname PYG@tok@nd\endcsname{\def\PYG@tc##1{\textcolor[rgb]{0.67,0.13,1.00}{##1}}}
\expandafter\def\csname PYG@tok@ne\endcsname{\let\PYG@bf=\textbf\def\PYG@tc##1{\textcolor[rgb]{0.82,0.25,0.23}{##1}}}
\expandafter\def\csname PYG@tok@nf\endcsname{\def\PYG@tc##1{\textcolor[rgb]{0.00,0.00,1.00}{##1}}}
\expandafter\def\csname PYG@tok@si\endcsname{\let\PYG@bf=\textbf\def\PYG@tc##1{\textcolor[rgb]{0.73,0.40,0.53}{##1}}}
\expandafter\def\csname PYG@tok@s2\endcsname{\def\PYG@tc##1{\textcolor[rgb]{0.73,0.13,0.13}{##1}}}
\expandafter\def\csname PYG@tok@vi\endcsname{\def\PYG@tc##1{\textcolor[rgb]{0.10,0.09,0.49}{##1}}}
\expandafter\def\csname PYG@tok@nt\endcsname{\let\PYG@bf=\textbf\def\PYG@tc##1{\textcolor[rgb]{0.00,0.50,0.00}{##1}}}
\expandafter\def\csname PYG@tok@nv\endcsname{\def\PYG@tc##1{\textcolor[rgb]{0.10,0.09,0.49}{##1}}}
\expandafter\def\csname PYG@tok@s1\endcsname{\def\PYG@tc##1{\textcolor[rgb]{0.73,0.13,0.13}{##1}}}
\expandafter\def\csname PYG@tok@sh\endcsname{\def\PYG@tc##1{\textcolor[rgb]{0.73,0.13,0.13}{##1}}}
\expandafter\def\csname PYG@tok@sc\endcsname{\def\PYG@tc##1{\textcolor[rgb]{0.73,0.13,0.13}{##1}}}
\expandafter\def\csname PYG@tok@sx\endcsname{\def\PYG@tc##1{\textcolor[rgb]{0.00,0.50,0.00}{##1}}}
\expandafter\def\csname PYG@tok@bp\endcsname{\def\PYG@tc##1{\textcolor[rgb]{0.00,0.50,0.00}{##1}}}
\expandafter\def\csname PYG@tok@c1\endcsname{\let\PYG@it=\textit\def\PYG@tc##1{\textcolor[rgb]{0.25,0.50,0.50}{##1}}}
\expandafter\def\csname PYG@tok@kc\endcsname{\let\PYG@bf=\textbf\def\PYG@tc##1{\textcolor[rgb]{0.00,0.50,0.00}{##1}}}
\expandafter\def\csname PYG@tok@c\endcsname{\let\PYG@it=\textit\def\PYG@tc##1{\textcolor[rgb]{0.25,0.50,0.50}{##1}}}
\expandafter\def\csname PYG@tok@mf\endcsname{\def\PYG@tc##1{\textcolor[rgb]{0.40,0.40,0.40}{##1}}}
\expandafter\def\csname PYG@tok@err\endcsname{\def\PYG@bc##1{\setlength{\fboxsep}{0pt}\fcolorbox[rgb]{1.00,0.00,0.00}{1,1,1}{\strut ##1}}}
\expandafter\def\csname PYG@tok@kd\endcsname{\let\PYG@bf=\textbf\def\PYG@tc##1{\textcolor[rgb]{0.00,0.50,0.00}{##1}}}
\expandafter\def\csname PYG@tok@ss\endcsname{\def\PYG@tc##1{\textcolor[rgb]{0.10,0.09,0.49}{##1}}}
\expandafter\def\csname PYG@tok@sr\endcsname{\def\PYG@tc##1{\textcolor[rgb]{0.73,0.40,0.53}{##1}}}
\expandafter\def\csname PYG@tok@mo\endcsname{\def\PYG@tc##1{\textcolor[rgb]{0.40,0.40,0.40}{##1}}}
\expandafter\def\csname PYG@tok@kn\endcsname{\let\PYG@bf=\textbf\def\PYG@tc##1{\textcolor[rgb]{0.00,0.50,0.00}{##1}}}
\expandafter\def\csname PYG@tok@mi\endcsname{\def\PYG@tc##1{\textcolor[rgb]{0.40,0.40,0.40}{##1}}}
\expandafter\def\csname PYG@tok@gp\endcsname{\let\PYG@bf=\textbf\def\PYG@tc##1{\textcolor[rgb]{0.00,0.00,0.50}{##1}}}
\expandafter\def\csname PYG@tok@o\endcsname{\def\PYG@tc##1{\textcolor[rgb]{0.40,0.40,0.40}{##1}}}
\expandafter\def\csname PYG@tok@kr\endcsname{\let\PYG@bf=\textbf\def\PYG@tc##1{\textcolor[rgb]{0.00,0.50,0.00}{##1}}}
\expandafter\def\csname PYG@tok@s\endcsname{\def\PYG@tc##1{\textcolor[rgb]{0.73,0.13,0.13}{##1}}}
\expandafter\def\csname PYG@tok@kp\endcsname{\def\PYG@tc##1{\textcolor[rgb]{0.00,0.50,0.00}{##1}}}
\expandafter\def\csname PYG@tok@w\endcsname{\def\PYG@tc##1{\textcolor[rgb]{0.73,0.73,0.73}{##1}}}
\expandafter\def\csname PYG@tok@kt\endcsname{\def\PYG@tc##1{\textcolor[rgb]{0.69,0.00,0.25}{##1}}}
\expandafter\def\csname PYG@tok@ow\endcsname{\let\PYG@bf=\textbf\def\PYG@tc##1{\textcolor[rgb]{0.67,0.13,1.00}{##1}}}
\expandafter\def\csname PYG@tok@sb\endcsname{\def\PYG@tc##1{\textcolor[rgb]{0.73,0.13,0.13}{##1}}}
\expandafter\def\csname PYG@tok@k\endcsname{\let\PYG@bf=\textbf\def\PYG@tc##1{\textcolor[rgb]{0.00,0.50,0.00}{##1}}}
\expandafter\def\csname PYG@tok@se\endcsname{\let\PYG@bf=\textbf\def\PYG@tc##1{\textcolor[rgb]{0.73,0.40,0.13}{##1}}}
\expandafter\def\csname PYG@tok@sd\endcsname{\let\PYG@it=\textit\def\PYG@tc##1{\textcolor[rgb]{0.73,0.13,0.13}{##1}}}

\makeatother

\DeclareMathAlphabet{\mathpzc}{OT1}{pzc}{m}{it}

\DeclareUnicodeCharacter{21D2}{$\Rightarrow$}
\DeclareUnicodeCharacter{2113}{$\mathpzc{l}$}
\DeclareUnicodeCharacter{2248}{$\approx$}
\DeclareUnicodeCharacter{2200}{$\forall$}
\DeclareUnicodeCharacter{2218}{$\circ$}
\DeclareUnicodeCharacter{2294}{$\sqcup$}
\DeclareUnicodeCharacter{3B1}{$\alpha$}
\DeclareUnicodeCharacter{3B2}{$\beta$}
\DeclareUnicodeCharacter{3B7}{$\eta$}
\DeclareUnicodeCharacter{3BB}{$\lambda$}
\DeclareUnicodeCharacter{3BC}{$\mu$}
\DeclareUnicodeCharacter{2080}{$_0$}
\DeclareUnicodeCharacter{2081}{$_1$}
\DeclareUnicodeCharacter{2082}{$_2$}
\DeclareUnicodeCharacter{2261}{$\equiv$}
\DeclareUnicodeCharacter{2E1}{$^l$}
\DeclareUnicodeCharacter{2B3}{$^r$}
\DeclareUnicodeCharacter{2032}{$'$}
\DeclareUnicodeCharacter{21D4}{$\Leftrightarrow$}
\DeclareUnicodeCharacter{27E8}{$\langle$}
\DeclareUnicodeCharacter{27E9}{$\rangle$}
\DeclareUnicodeCharacter{27F6}{$\longrightarrow$}
\DeclareUnicodeCharacter{220E}{$\blacksquare$}
\DeclareUnicodeCharacter{27FA}{$\Leftrightarrow$}
\DeclareUnicodeCharacter{22A4}{$\top$}
\DeclareUnicodeCharacter{22A5}{$\bot$}
\DeclareUnicodeCharacter{03A3}{$\Sigma$}
\DeclareUnicodeCharacter{03C0}{$\pi$}

\makeatletter
\AtBeginEnvironment{minted}{\dontdofcolorbox}
\def\dontdofcolorbox{\renewcommand\fcolorbox[4][]{##4}}
\makeatother

%% file: macros.tex
\newcommand{\opt}[1]{\texttt{-{}-#1}}

\newcommand{\catC}{\ensuremath{\mathcal{C}}}

\newcommand{\mor}[2]{\ensuremath{#1 \Rightarrow #2}}
\newcommand{\comp}[2]{\ensuremath{#1 \circ #2}}

\newcommand{\hcheckmark}{\checkmark\kern-1.1ex\raisebox{.7ex}{\rotatebox[origin=c]{125}{\fontseries{bx}\selectfont
--}}}

\newcommand{\xmark}{\ding{55}}

\newcommand\agda[1]{\mintinline{agda}{#1}}

\newcommand{\catset}{\mathpzc{Set}}

%% file: formal.tex
\section{Design Choices}\label{sec:basics}

Choices arise from both the system and its logic, as well as from the domain
itself.

\subsection{Fitting with Agda}\label{sec:agda}

The previous formalization~\cite{copumpkin} was done in a much older Agda,
with a seriously under-developed standard library. To better fit with
\emph{modern} Agda, we choose to:
\begin{enumerate}
\item use dependent types,
\item be constructive,
\item re-use as much of the standard library~\cite{agda-stdlib} as possible,
\item use the naming convention of its standard library whenever meaningful,
\item use the \texttt{variable} generalization feature for levels and categories,
\item try to fit with as many modes of Agda as possible.
\end{enumerate}

The first two requirements are natural, as choosing otherwise would create
a clash of philosophy between the system and one of its libraries.
The next two are just good software engineering, while the fifth is mere
convenience.  Note that re-using the standard library pushes us towards
setoids (more on that later) as its formalization of algebra uses them
extensively.

The last requirement is more subtle: we want to allow others to use alternative
systems or make postulates if
they wish, and still be able to use our library.  This means that we need to avoid
using features that are incompatible with supported systems in Agda.  For example, when
added to Martin-L\"{o}f Type Theory (MLTT)~\cite{DBLP:books/daglib/0000395}, axiom
K~\cite{streicher1993investigations}, equivalent to \emph{Uniqueness of Identity
  Proofs} (UIP), creates a propositionally extensional type theory incompatible with
\emph{univalence}~\cite{hottbook}.  Thus Agda has options such as
\opt{without-K}~\cite{10.1145/2628136.2628139} to access the intensional type theory
MLTT, and conversely \opt{with-K} to turn on axiom K.  Separately, there is cubical
type theory (\opt{cubical})~\cite{Vezzosi:2019:CAD:3352468.3341691} which implements a
computational interpretation of homotopy type theory (HoTT)~\cite{hottbook} and supports univalence.  Intensional type
theory is compatible with both options of \opt{with-K} and \opt{cubical}, and thus if we build our library using
\opt{without-K}, it can be maximally re-used.  This further implies that we have to
avoid propositional equality as much as possible, as pure MLTT gives us very few tools to
work with it. We additionally turn on the \opt{safe} option to avoid possible misuses
of certain features which could lead to logical inconsistencies. 

\subsection{Which Category Theory?}\label{sec:ct}

Category theory
is often presented as a single theory, but there are in fact a
wealth of flavours: set-theoretic, where a
category has a single hom-set equipped with source and target maps;
ETCS-style~\cite{10.2307/72513}, where there are no objects at all; dependently-typed, where
hom-``sets'' are parametrized by two objects; proof-irrelevant, where the
associativity and identity laws are considered to be
unique~\cite{ahrens_kapulkin_shulman_2015,UniMath,10.1007/978-3-319-08970-6_18,copumpkin};
setoid-based, where each category relies on a local notion of equivalence of
hom-sets rather than relying on a \emph{global} equality
relation~\cite{coq:category,copumpkin}.  There are also questions of being strict
or weak, whether to do $1$-categories, $n$-categories or even
$\infty$-categories.  What to choose?

Standard textbooks often define a category as follows:

\begin{definition}\label{def:cat}
  A category \catC\xspace consists of the following data:
  \begin{enumerate}
  \item a collection of objects, $\catC_0$,
  \item a collection of morphisms, $\catC_1$, between two objects. We use $f :
    \mor{A}{B}$ to denote the morphism $f
    \in \catC_1$ is
    between objects $A$ and $B$,
  \item for each object $A$, we have an identity morphism $1_A :
    \mor{A}{A}$, and
  \item morphism composition $\circ$ composing two morphisms $f : \mor{B}{C}$ and $g :
    \mor{A}{B}$ into another morphism $\comp f g : \mor{A}{C}$.
  \end{enumerate}
  These must satisfy the following laws:
  \begin{enumerate}
  \item identity: for any morphism $f : \mor A B$, we have $\comp f {1_A} = f = \comp
    {1_B} f$, and
  \item associativity: for any three morphisms $f$, $g$ and $h$, we have $\comp {(\comp
    f g)} h = \comp f {(\comp g h)}$.
  \end{enumerate}
\end{definition}

Embedded in the above definition are a variety of decisions, and we will use
these as a running example to explain ours.

\subsubsection{Collections}\label{sec:collections}
The first item to notice is the use of \emph{collection} rather than
\emph{set} or \emph{type}. Textbooks tend to do this to side-step
``size'' issues, and then define various kinds of categories depending
on whether each of the collections (objects, all morphisms, all morphisms
given a pair of objects) is ``small'', i.e. a set.  This matters because a
number of constructions in category theory produce large results. 

We define \emph{collections of objects} to be \emph{types}, with no further
assumptions or requirements.  We do know that in MLTT types are well modeled by
$\infty$-groupoids~\cite{Hofmann96thegroupoid,warren2008homotopy} --- so
wouldn't this
higher structure be a problem? No! This is because we never look at it, i.e.
we never look at the identity type (or their identity types) of objects.

The collection of morphisms is trickier, and splits into:
\begin{enumerate}
\item Is there a single collection of morphisms?
\item What about equality of morphisms?
\end{enumerate}
The first item will be treated here, the second in subsection~\ref{sec:proof-rel}.

If we try to put all the morphisms of certain categories together in a single
collection, size issues arise, but there is also another issue: if we consider
composition as a function of pairs of morphisms, then this function is partial.
Luckily, our dependent type theory allows one to side-step both issues at the same
time: rather than a single collection of morphisms, we have a (dependently-typed)
family of morphisms, one for each pair of objects. In category theory, one rarely considers
the ``complete collection'' of all morphisms.  This solves the composition
problem too, as we can only compose morphisms that have the right type, leading
to the following (partial) definition:
\begin{minted}{agda}
record Category : Set where
  field
    Obj : Set
    _⇒_ : (A B : Obj) → Set
    _∘_ : ∀ {A B C} → B ⇒ C → A ⇒ B → A ⇒ C
\end{minted}

\subsubsection{Strictness}\label{sec:strictness}

Traditional textbooks tend to implicitly assume that collections are somewhat
still set-like, in that \emph{equality} is taken for granted, i.e. that
it always makes sense to ask whether two items from a collection are
equal. Not just that it always makes sense, but that the underlying
meta-theory will always answer such queries in finite time%
\footnote{That we should not ask whether two objects are equal is an issue
well described at the \emph{Principle of Equivalence} page of the nLab.
\url{https://ncatlab.org/nlab/show/principle+of+equivalence}}.

The \emph{Principle of Isomorphism}~\cite{makkai1998towards} already tells us that we
should not assume that we have any relation on objects other than the one given by
categorical principles (isomorphism); a related \emph{Principle of Equivalence}%
~\cite{Ahrens2019} can be stated formally in the context of homotopy type
theory.  That we normally do not have, and should not assume, such a relation
have motivated some to create the concept of a \emph{strict category}, where
we have given ourselves the ability to compare objects for equality.  Classically,
sets have equality defined as a total relation, so that this comes ``for free''.  In
other words, given two elements $x,y$ of a set $S$, in set theory it always makes
sense to ``ask'' the question $x = y$, and this has a \emph{boolean} answer.  This is
one reason why it took a while for the Principle of Equivalence to emerge as
meaningful.  As global extensionality is hard to mechanize in MLTT, it is simplest to forgo having an
equality relation on objects at all.

\subsubsection{Proof-relevant Setoids}\label{sec:proof-rel}

In \Cref{def:cat}, equality of morphisms is also taken for granted.  The
laws use equality, blithely assume that the meta-theory defines it.  In
MLTT, which equality we use matters.
Usually, there are three options: local equality (setoids), propositional
equality in intensional type theory (\texttt{\_≡\_}),
and propositional equality with UIP.

Propositional equality does not work very well in MLTT without further
properties or axioms to deal with functions (e.g. function extensionality),
while many categories have (structured) functions as
morphisms. The third case is a plausible option, because UIP relieves
us from reasoning about equality between equalities due to UIP and reduces the
issue to familiar set theory. Nonetheless, the \opt{with-K} mode and the \opt{cubical} mode
approach  UIP in different ways and it is not immediate to us how to organize the
library so that it is compatible with both. Thus this option, though very
interesting, seems to clash with our original motivation. 

For these reasons, we chose to work with setoids. Earlier formalizations of category theory in type theory already used
setoids~\cite{aczel1993galois,huet2000constructive,copumpkin,coq:category}, which associate an
equivalence relation to each type.  This generalizes ``hom-sets'' to
``hom-setoids'',  i.e. the definition of
category is augmented as follows:
\begin{minted}{agda}
  _≈_   : ∀ {A B} → (f g : A ⇒ B) → Set
  equiv : ∀ {A B} → IsEquivalence (_≈_ {A} {B})
\end{minted}
In both types, \texttt{A} and \texttt{B} are objects in the current
category. \texttt{IsEquivalence} is a predicate provided by the standard library that expresses
that \texttt{\_≈\_} is an equivalence relation. Furthermore, composition must respect
this equivalence relation, which we can express as%
\footnote{We use variable generalization to leave implicit variables out and let Agda
  infer them, so we will omit unnecessary type ascriptions provided an unambiguous context..}:
\begin{minted}{agda}
  ∘-resp-≈  : f ≈ h → g ≈ i → f ∘ g ≈ h ∘ i
\end{minted}
Note that
\texttt{\_≈\_} can be specialized to 
\texttt{\_≡\_} to work in other settings such as cubical type theory.

We explicitly do not assume that two witnesses of \linebreak \agda{_≈_}
are equivalent, making our setoids \emph{proof relevant}.
Proof-relevance is a significant difference between this library and the
previous one~\cite{copumpkin},  which relied heavily on irrelevant
arguments~\cite{lmcs:1045}.  In particular, all of the proof obligations
(for example left and right identities, and associativity in the case of a category)
were marked irrelevant in~\cite{copumpkin}, making these proofs ``unique'' by fiat.  Thus
two categories that differed only in their proofs
were automatically regarded as (definitionally) equal.  Ignoring the
details of proofs is convenient --- but unfortunately irrelevant arguments are not part
of MLTT. Worse yet, they are not a stable, well-maintained feature in Agda, so we
refrained from using this feature in our library.

We gain other improvements over the previous library by having hom-setoids 
proof-relevant. In \cite{copumpkin}, due to irrelevance,
the content of \texttt{\_≈\_} is ignored. However, this is not necessarily coherent
under all settings. For example, when defining the (large) category of all categories,
with proof relevance, we can use natural isomorphisms as equivalence between functors.
In other words, in our setting, the ``natural'' definition of the (large) category
of all categories is a category, we do not need to move up to $2-$categories.
The previous library, contrarily, must use heterogeneous
equality for equivalence between functors, which subsequently required axiom K for
elimination and restricted the possible choice of foundations. In this case, making setoids proof-relevant actually allowed us to
internalize more category theory into itself.

Libraries formalizing category theory based on
HoTT~\cite{ahrens_kapulkin_shulman_2015,10.1007/978-3-319-08970-6_18} restricts
hom-sets to be hSets by requiring an additional law which states the contractibility of
equality proofs between equalities in hom-sets.  Our library implements a settings
which allow richer structures in the hom-setoids.

\subsubsection{Explicit Universe Level}\label{sec:ulevel}

In Agda, users are exposed to the explicit handling of universe levels
(i.e. of type \texttt{Level}).
Some find it cumbersome, but we have found it quite useful.  To help with
reuse, we make our definitions universe-polymorphic by parameterizing
them by \texttt{Level}s. For example, a \texttt{Category} is refined as follows:
\begin{minted}{agda}
record Category (o ℓ e : Level)
    : Set (suc (o ⊔ ℓ ⊔ e)) where
  field
    Obj : Set o
    _⇒_ : (A B : Obj) → Set ℓ
    _≈_ : ∀ {A B} → (f g : A ⇒ B) → Set e
  -- other fields omitted
\end{minted}
Since the definition of \texttt{Category} contains three \texttt{Set}s representing
objects, morphisms and the equivalence relations respectively, it can be
indexed by three \texttt{Level}s and thus live at least one level above their supremum.

One significant advantage of a level-parametric definition is that it
simplifies the formalization of concepts such as the category of categories, or
that of functors.  We do not have to duplicate definitions, nor do we
have to sprinkle various size constraints about (such as a category being
``locally small'') to avoid set-theoretic troubles.

With explicit \texttt{Level}s, new phenomena become visible. In set-based category
theory, one might be tempted to talk about the (large) category of all sets or
all setoids.  In Agda, we can only talk about the category of all
\texttt{Setoid}s with particular \texttt{Level}s:
\begin{minted}{agda}
Setoids : ∀ c ℓ →
  Category (suc (c ⊔ ℓ)) (c ⊔ ℓ) (c ⊔ ℓ)
Setoids c ℓ = record
  { Obj       = Setoid c ℓ
    -- ... other fields omitted.
  }
\end{minted}
Here \texttt{c} and \texttt{ℓ} are the \texttt{Level}s of the carrier
and the equivalence of a \texttt{Setoid c ℓ}, respectively. We can clearly see the
ensuing size issue.  The definition \emph{must} be indexed by \texttt{Level}s,
as there is no term \emph{in} the type theory in which all \texttt{Setoid}s (for example)
exist.  The set of types \texttt{Set ℓ} is somewhat analogous to a
\emph{Grothendieck universe} which provides a way to resolve Russell-style paradox in
set theory, as it is closed under similar operations, but not
unrestricted unions, where one must then move to a larger universe.
\texttt{Set (suc ℓ)} is indeed sometimes called a Russell-style universe.

However, universes in Agda are non-cumulative by default.  Combined with
explicit \texttt{Level}s, this leads to other issues. With
cumulative universes, a type in one universe automatically inhabits all larger
universes. In Agda, one must explicitly lift terms to larger
levels, which adds a certain amount of ``noise'' to some code.  For example, consider
two categories of \texttt{Setoid}s, \texttt{Setoids 0 1} and \texttt{Setoids 1 1},
differing only in their first indices.  With cumulative universes,
even though we still need
to apply a lifting functor to embed \texttt{Setoids 0 1} in \texttt{Setoids 1 1}, the
functor is trivially defined:
\begin{minted}{agda}
liftF = record  -- we are defining a functor
  { F₀ = λ x → x
    -- other fields are omitted
  }
\end{minted}
With cumulativity, the second \texttt{x} has a larger universe than the
first one. Without cumulativity, explicit calls to \texttt{lift} must be inserted:
\begin{minted}{agda}
liftF = record  -- we are defining a functor
  { F₀ = λ x → lift x
    -- other fields are omitted
  }
\end{minted}
We noticed that when handling some classical definitions or results involving sets, like
adjoint functors and the Yoneda lemma, we often need to postcompose with a lifting
functor in order to achieve the most general statements. For
example, the Yoneda lemma involves the natural isomorphism in $X$:
\begin{align*}
  Nat[yX, F] \simeq FX
\end{align*}
where $F : \mathcal{C}^{op} \Rightarrow Set$ for some category $\mathcal{C}$ and
$X \in \mathcal{C}$ is an object. In the actual formalization, assuming $\mathcal{C}$
has type \texttt{Category o ℓ e} and $F$ maps to \texttt{Setoids ℓ e}, then by some
calculation, we see that $Nat[yX, F]$ actually maps to \texttt{Setoids (o ⊔ ℓ ⊔
  e) (o ⊔ ℓ ⊔ e)}, because the \texttt{Setoids} must be large enough to contain
$F$. Thus we cannot create this natural isomorphism without
lifting the universe on the right hand side to the correct level.
Explicit universe \texttt{lift}ing and \texttt{lower}ing are then
required in subsequent equational reasoning, which quickly become rather annoying.

Since 2.6.1, Agda has an experimental feature of cumulative universes. We hope
that this feature may help us remove some clutter in our statements and
proofs. However, at present, cumulativity is not deemed
\opt{safe}. Furthermore, we encountered issues with the level constraint solver when
we experimented with adapting our library to that environment.

\subsection{Duality}\label{sec:duality}

In category theory, duality is omnipresent. However, in type theory and
in formalized mathematics, subtleties arise.  Some are due to proof relevance,
while others are usability issues, which we discuss here.

\paragraph{Additional Laws for Duality}

In category theory, there is a very precise sense in which, if a theorem holds,
then its dual statement also holds. Thus, in theory, we obtain two theorems by
proving one. This is the \emph{Principle of
Duality}~\cite{Awodey:2010:CT:2060081}, which we would like to exploit.

But first, we need to make sure that the most basic duality, that of forming
the opposite category, should be involutive. We can easily prove that the
double-opposite of a category $C$ is equivalent to $C$. This equivalence
is true \emph{definitionally} with proof-irrelevant definitions in~\cite{copumpkin}.  Can we recover
this here as well?  Yes -- we can follow~\cite{10.1007/978-3-319-08970-6_18} and require two (symmetric) proofs of
associativity of composition in the definition of a \texttt{Category}:
\begin{minted}{agda}
    assoc     : (h ∘ g) ∘ f ≈ h ∘ (g ∘ f)
    sym-assoc : h ∘ (g ∘ f) ≈ (h ∘ g) ∘ f
\end{minted}
Specifically, with \texttt{sym-assoc}, we can define its opposite category as follows:
\begin{minted}{agda}
  op : Category o ℓ e
  op = record
    { assoc     = sym-assoc
    ; sym-assoc = assoc
    -- other fields omitted
    }
\end{minted}
Otherwise, without \texttt{sym-assoc}, we would have to use the symmetry
of \texttt{\_≈\_}:
\begin{minted}{agda}
  assoc = sym assoc
\end{minted}
But now, applying duality twice gives \texttt{sym (sym assoc)} for the
associativity proof, which is not definitionally
equal to \texttt{assoc}.  This makes the properties of an opposite category less
useful than ones of the original one. For example, we might want to prove some
properties about coproducts by proving the dual properties about products in the
opposite category. Without involution of \texttt{op}, we would have to argue the
properties still hold if we swap to another associativity proof, which defeats the
usefulness of the Principle of Duality.

Another convenient law to add is
\begin{minted}{agda}
    identity² : id ∘ id ≈ id
\end{minted}
This law can be proved by taking \texttt{f} as \texttt{id} in either the left
identity or right identity law:
\begin{minted}{agda}
    identityˡ : id ∘ f ≈ f
    identityʳ : f ∘ id ≈ f
\end{minted}
We add this additional law for the following reasons:
\begin{enumerate}
\item When proving \texttt{id ∘ id ≈ id}, we need to choose between \texttt{identityˡ}
  and \texttt{identityʳ}, while there is no particular reason to prefer one to
  another. Adding this law neutralizes the need to make this choice.
\item In the implementation, we sometimes rely on constant functors, which ignore the
  domain categories and constantly return fixed objects in the codomain categories and
  their identity morphisms. Since the domain categories are completely ignored, these functors
  are intuitively ``the same'' as their duals. \texttt{identity²} allows constant functors
  to be definitionally equal to their duals even with proof-relevance.
\end{enumerate}

\paragraph{Independent Definitions of Dual Concepts}

In other
libraries~\cite{timany_et_al:LIPIcs:2016:6000,UniMath,ahrens_kapulkin_shulman_2015,10.1007/978-3-319-08970-6_18},
it is typical to define one concept and use duality to obtain the opposite one. For
example, we could define the initial object of $\mathcal{C}$, \texttt{Initial
  $\mathcal{C}$} as usual, and then define the terminal object by taking the opposite
as follows:
\begin{minted}{agda}
  Terminal′ : ∀ {o ℓ e} (C : Category o ℓ e) →
               Set _
  Terminal′ C = Initial (Category.op C)
\end{minted}

However, we do not take this approach. Instead, we define concepts
explicitly in terms of
data and laws and define conversions between duals in modules of the form
\texttt{*.Duality}. This has the following advantages:
\begin{enumerate}
\item when constructing or using the concepts, the names of the fields are more
  familiar;
\item theorems relating redundant definitions increase our confidence that our
  definitions are correctly formulated;
\item the redundancy helps maintain the Principle of Duality.
\end{enumerate}

Expanding on this third point: like with \texttt{sym-assoc},
we want duality to be a definitional involution for a number of concepts.  We were able to
identify a number of concepts which require additional laws to achieve this goal,
which we detail next.

\paragraph{Duality-Completeness of Laws}

Ensuring the
involution of duality turns out to be a very general design principle. We sometimes
obtain it for free, e.g. \texttt{Functor} and \texttt{Adjoint}. In other cases, we
need to supply a symmetric version of a law. For example, \texttt{Category},
\texttt{NaturalTransformation}, \linebreak \agda{Dinatural} (transformation) and
\texttt{Monad} all need some extra laws. As a rule of thumb,
if a conversion to the dual concept requires equational reasoning, even
as simple as applying \texttt{sym} to \texttt{assoc},
then we need to add that equation as a law.  In other words, our laws should
either be self-dual, or come in dual pairs (quite reminiscent of work on
reversible computation~\cite{carette2016groupoids} where the same property is
desirable). 
We ensure this principle by proving theorems of the
following form:
\begin{minted}{agda}
op-involutive : Category.op C.op ≡ C
op-involutive = ≡.refl
\end{minted}
Here \texttt{C} is a \texttt{Category}. We also supply similar proofs for conversions
between dual concepts, e.g.:
\begin{minted}{agda}
op⊤⟺⊥ : (⊥ : Initial) →
  op⊤⇒⊥ (⊥⇒op⊤ ⊥) ≡ ⊥
op⊤⟺⊥ _ = ≡.refl
\end{minted}
\texttt{⊥⇒op⊤} converts an initial object to a terminal object in the opposite
category and \texttt{op⊤⇒⊥} does the inverse. 
We put these theorems in \texttt{private}
blocks so they are only type checked. These theorems must be proved precisely by reflexivity. This ensures
that our definitions are duality-friendly.

Once we get the definition right, we also provide a helper constructor without the
additional laws, so that defining these self-dual versions are not more cumbersome
than their classical counterpart. Constructions defined through the helpers still enjoy the principle
of duality. Consider an application of the helper for \agda{Category}, which
effectively proves \texttt{sym-assoc} by applying symmetricity:
\begin{minted}{agda}
Some-Cat = record {
  -- other fields ignored
  assoc = some-proof ; sym-assoc = sym some-proof
}
\end{minted}
Notice that \texttt{Category.op (Category.op Some-Cat)} remains definitionally equal
to \texttt{Some-Cat}. 
In general, we found that the addition of these extra laws were
beneficial in the setting of 1-category theory.  The situation becomes more complex when we
move to the \texttt{Bicategory} setting, as we
must consider higher structures. 
Exactly how to modify the definitions of higher structures to obtain similar
good behaviour with respect to definitional equalities is left as future work.

\subsection{Encodings as Records}\label{sec:encap}

Another important design decision is how to encode definitions. Generally,
two different styles are used:
records~\cite{coq:category,10.1007/978-3-319-08970-6_18} or nested $\Sigma$
types~\cite{ahrens_kapulkin_shulman_2015,UniMath}. In the latter style,
developers typically need to write a certain amount of boilerplate accessor
code. In Agda it is more natural to use record definitions:
\begin{enumerate}
\item It aligns very well with the design principle of the standard library,
\item Records allow various \emph{syntactic sugar}, as well as having good
  IDE (via Emacs) support,
\item Most importantly records also behave as modules. That is, we can
  export symbols to the current context from a record when it is unambiguous to do
  so.
\end{enumerate}

The record module feature enables some structural benefits as well.
Consider the following definition of a \texttt{Monad} over a category:
\begin{minted}{agda}
record Monad {o ℓ e} (C : Category o ℓ e)
    : Set (o ⊔ ℓ ⊔ e) where
  field
    F : Endofunctor C
    η : NaturalTransformation idF F
    μ : NaturalTransformation (F ∘F F) F
  -- ... laws are omitted
\end{minted}

We often need to refer to components of the \texttt{Functor} \texttt{F} or the
\texttt{NaturalTransformation}s $\eta$ or $\mu$ when working with a \texttt{Monad}. By
adding the following module definitions to the \texttt{Monad} record, we can use
dot accessors to access deeper fields:
\begin{minted}{agda}
  module F = Functor F
  module η = NaturalTransformation η
  module μ = NaturalTransformation μ
\end{minted}
For example, if we have two \texttt{Monad}s \texttt{M} and \texttt{N} in scope, we
can declare \texttt{module M = Monad M} and \texttt{module N = Monad N}, and get
the following convenient nested dot accessors:
\begin{minted}{agda}
  M.F.₀  -- the mapping of objects of F of M
  N.F.₁  -- the mapping of morphisms of F of N
  M.μ.η X
    -- the component of the NaturalTransformation
    -- μ of M at object X
  N.η.commute f  -- the naturality square of the
                 -- NaturalTransformation η of N
                 -- at morphism f
\end{minted}
The original syntax is more verbose, so the module syntax is
significantly more convenient:
\begin{minted}{agda}
  Functor.F₀ (Monad.F M)
  Functor.F₁ (Monad.F N)
  NaturalTransformation.η (Monad.μ M) X
  NaturalTransformation.commute (Monad.η N) f
\end{minted}
Another frequent style is to open a module with renaming:
\begin{minted}{agda}
  open NaturalTransformation (Monad.μ M)
    renaming (η to α)
  open NaturalTransformation (Monad.η N)
    renaming (η to β)
\end{minted}
Then we use $\alpha$ and $\beta$ to refer to the component maps of the corresponding natural
transformations. Unfortunately such setup code is ad-hoc and inconsistent
across files.

We use the accessor module style throughout the code base, as it feels more elegant
and readable to us than other styles.

%% file: method.tex
\section{Formalization and Definitions}\label{sec:pr-def}

While implementing the library, we noticed several times that ``standard''
definitions needed to be adjusted, for technical reasons. Certain direct translations
of concepts from classical category theory are not even well-typed!
Proof-relevance also forces us to pay close attention to the laws embedded in
each concept, to obtain more definitional equalities, rather than relying on
extensional behavior for ``sameness''.  The resulting formalization is more
robust, and it also eases type checking.

Various categorical concepts are well-known to have multiple, equivalent definitions.
We have found that, although classically equivalent, some turn out to be technically
superior for our formalization. We are sometimes even forced to introduce new ones.
Here we discuss the choices we made when defining concepts related to closed monoidal
categories and finite categories in detail, focusing on the underlying rationale.

\subsection{Adjoint Functors}\label{sec:adjoint}

Adjoint functors are frequently regarded as one of the most fundamental concepts
in category theory and play a critical part in the definition of closed
monoidal categories.  The following two definitions of adjoint functors are
equivalent in classical category theory.

\begin{definition}\label{def:adj:1}
  Functors $F : \mathcal{C} \Rightarrow \mathcal{D}$ and
  $G : \mathcal{D} \Rightarrow \mathcal{C}$ are adjoint, $F \dashv G$, if there is a
  natural isomorphism $Hom(F X, Y) \simeq Hom(X, G Y)$ in $X$ and $Y$.
\end{definition}

\begin{definition}\label{def:adj:2}
  Functors $F : \mathcal{C} \Rightarrow \mathcal{D}$ and
  $G : \mathcal{D} \Rightarrow \mathcal{C}$ are adjoint, $F \dashv G$, if there
  exist two natural transformation, unit $\eta : 1_{\mathcal{C}} \Rightarrow G F$ and counit
  $\epsilon : F G \Rightarrow 1_{\mathcal{D}}$, so that the
  triangle identities below hold:
  \begin{enumerate}
  \item $\epsilon F \circ F \eta = 1_F$
  \item $G \epsilon \circ \eta G = 1_G$
  \end{enumerate}
\end{definition}

These two definitions are classically equivalent. 
\Cref{def:adj:1} is typically very easy to use in classical category theory, as it
it is about \emph{hom-sets}, and so partly set-theoretic in its formulation.
However, this definition is not natural in Agda, especially in the presence
of non-cumulative universes and level-polymorphic morphisms (\Cref{sec:ulevel}),
so that the morphisms of $\mathcal{C}$ and $\mathcal{D}$ do not always
live in the same universe level. Thus $Hom(FX, Y) \simeq Hom(X, GY)$ is not
well-typed as is.  Instead, $Hom(FX, Y)$ and $Hom(X, GY)$
need to be precomposed by lifting functors, which lift both hom-setoids to the
universe at their supremum level.  One might think that this technicality is
classically not present -- but that is because many textbooks make the blanket
assumption that all their categories are locally small. It corresponds to assuming
that the morphisms of $\mathcal{C}$ and $\mathcal{D}$ live at the same
(lowest!) universe level. In that case, we indeed do not need the lifting
functors. This ``technical noise'' add by the lifts get rid of this problem,
but set theory has no means
to express size polymorphism (as in set, proper class, superclass, etc).
However, such coercions are neither intuitive nor easy to work with.

\Cref{def:adj:2}, on the other hand, has no such problem. Both natural transformations
and triangle identities involve no explicit universe level management.  For this
reason, we choose \Cref{def:adj:2} as our primary definition of adjoint functors and
have \Cref{def:adj:1} as a theorem. The added
polymorphism of the unit-counit definition makes it more suitable when working in type
theory.

\subsection{Monoidal Category}

A monoidal category can be understood as a generalization of a monoid to the categorical
setting. Classically, a monoidal
category has the following definition~\cite{KELLY1964397}:
\begin{definition}\label{def:monoidal}
  A category $\catC$ is monoidal with the following data:
  \begin{enumerate}
  \item a unit object $u$, 
  \item a bifunctor $\otimes$, 
  \item for any object $X$, a natural isomorphism $\lambda$ of $u \otimes X \simeq X$,
  \item for any object $X$, a natural isomorphism $\rho$ of $X \otimes u \simeq X$, and
  \item for any objects $X$, $Y$ and $Z$, a natural isomorphism $\alpha$ of
    $(X \otimes Y) \otimes Z \simeq X \otimes (Y \otimes Z)$.
  \end{enumerate}
  They satisfy the following diagrams for any objects $X$, $Y$, $Z$ and $W$:
  
  \begin{center}
\begin{tikzcd}
(X \otimes u) \otimes Y \arrow[d, "\alpha"'] \arrow[r, "\rho \otimes 1_Y"] & X \otimes Y \\
X \otimes (u \otimes Y) \arrow[ru, "1_X \otimes \lambda"']                 &            
\end{tikzcd}

\begin{tikzcd}
(X \otimes Y) \otimes (Z \otimes W) \arrow[rd, "\alpha"]                                &                                                                      \\
((X \otimes Y) \otimes Z) \otimes W \arrow[d, "\alpha \otimes 1_W"] \arrow[u, "\alpha"] & X \otimes (Y \otimes (Z \otimes W))                                  \\
(X \otimes (Y \otimes Z)) \otimes W \arrow[r, "\alpha"]                                 & X \otimes ((Y \otimes Z) \otimes W) \arrow[u, "1_X \otimes \alpha"']
\end{tikzcd}
  \end{center}
\end{definition}

The associativity of the natural isomorphism
$\alpha$ is problematic as
$(X \otimes Y) \otimes Z$ has type \texttt{Functor
  $((\mathcal{C} \times \mathcal{C})\times \mathcal{C})$ $\mathcal{C}$}, while
$X \otimes (Y \otimes Z)$ has type \texttt{Functor
  $(\mathcal{C} \times (\mathcal{C}\times \mathcal{C}))$ $\mathcal{C}$}.  As the
domains are not definitionally equal, there cannot be a natural isomorphism
between them. For type correctness, one possible solution is to precompose the
first functor with an associator from
$(\mathcal{C} \times \mathcal{C})\times \mathcal{C}$ to
$\mathcal{C} \times (\mathcal{C}\times \mathcal{C})$.  This is not mere
pedantry: we know that ``one level up'', this is an unavoidable issue.  In other
words, some issues that show up as type-checking problems in 1-category theory are
actually previews of 2-categorical subtleties ``peeking through'', that can be ignored
in paper-math. Our definition instead asks for the following data:
\begin{enumerate}
\item an isomorphism between $(X \otimes Y) \otimes Z$ and $X \otimes
  (Y \otimes Z)$, for any objects $X$, $Y$ and $Z$, and
\item two naturality squares to complement the missing laws so that the isomorphism above
  is natural. 
\end{enumerate}
This leads to a definition that is easier to use, and the required natural
isomorphism becomes a theorem. 

\subsection{Closed Monoidal Category}

Intuitively, a closed monoidal category is a category possessing both a closed
and a monoidal structure, in a compatible way. In the literature, we can
find various definitions of a closed monoidal category:
\begin{enumerate}
\item\label{def:cmc:1} (a monoidal category with an added closed structure): given a
  monoidal category (with bifunctor $\otimes$), there is also
  a family of functors $[X, -]$ for each object $X$, such that
  $- \otimes X \dashv [X, -]$. The closed bifunctor (or inner hom) $[-, -]$ is then induced uniquely
  up to natural isomorphism.
\item (a closed category with an added monoidal structure): given a closed category with
  bifunctor $[-, -]$, it is additionally equipped with a family of functors
  $- \otimes X$ for each object $X$, such that $- \otimes X \dashv [X, -]$. The monoidal
  bifunctor $\otimes$ is then induced uniquely up to natural isomorphism.
\item\label{def:cmc:3} (via a natural isomorphism of hom-sets): given a
  category, for each object $X$, there are two families of functors $- \otimes X$ and
  $[X, -]$, such that the isomorphism $Hom(Y \otimes X, Z) \simeq Hom(Y, [X, Z])$ is
  natural in $X$, $Y$ and $Z$. Both bifunctors $\otimes$ and $[-, -]$ are then
  induced uniquely up to natural isomorphism.  
\end{enumerate}

Note that the third definition above is not biased towards either the closed or
monoidal structure.  All three can be shown equivalent (classically).  But in
the proof-relevant setting, problems arise.  One problem that all three
definitions share is that they all induce at least one bifunctor from a
\emph{family} of functors. For example, in the first definition, the closed
bifunctor $[-,-]$ is the result of a theorem; two different instances of
$[-,-]$ (which might potentially differ in their proofs) can only be
related by a natural isomorphism, which is often too weak. In other words,
we want \emph{both} bifunctors $\otimes$ and $[-, -]$ to be part of the definition so
that they can be constructed elsewhere and they are related by other laws.
None of the three definitions above satisy this requirement.  We thus arrive at the
following definition, which is the one we use:

\begin{definition}\label{def:cmc:4}
  A closed monoidal category is a category with two bifunctors $\otimes$ and
  $[-,-]$, so that
  \begin{enumerate}
  \item $\otimes$ satisfies the laws of a monoidal category,
  \item  $- \otimes X \dashv [X ,-]$ for each object $X$, and
  \item for a morphism $f : X \Rightarrow Y$, the induced natural transformations
    $\alpha_f : - \otimes X \Rightarrow - \otimes Y$ and
    $\beta_f : [Y, -] \Rightarrow [X, -]$ form a mate (or a conjugate in the sense
    of~\cite{maclane:71}) for the two adjunctions,
    $- \otimes X \dashv [X ,-]$ and $- \otimes Y \dashv [Y ,-]$, formed by previous
    constraint.
  \end{enumerate}
\end{definition}

This definition is \emph{better}, in the sense that it is 1) unbiased, 2) incremental
(it simply adds more constraints on both bifunctors). Further note that both
bifunctors are given as part of the data, rather than derived, which allows us to
consistently refer to both uniquely. The
following theorem strengthens our confidence:
\begin{theorem}
  A closed monoidal category according to~\Cref{def:cmc:4} is a closed category. 
\end{theorem}
In addition, the closed bifunctor $[-,-]$ from the closed category in this theorem is
definitionally the same one given in~\Cref{def:cmc:4}. This allows closed monoidal
categories to inherit all properties of closed categories as they are talking about
precisely the same $[-,-]$.

A potential downside of this definition is that it depends on \emph{mates} which are not
present in previous definitions.  Though this seems to add complexity, we
argue that the benefit is worth the effort. We now discuss mates in order to justify
that this new definition is equivalent to the previous three.

\subsection{Mate}\label{sec:mate}

Mates express naturality between adjunctions. They are typically defined by two
natural isomorphisms between hom-sets as follows:

\begin{definition}\label{def:mate:1}
  For functors $F, F' : \mathcal{C} \Rightarrow \mathcal{D}$ and
  $G, G' : \mathcal{D} \Rightarrow \mathcal{C}$, two natural transformations
  $\alpha : F \Rightarrow F'$ and $\beta : G' \Rightarrow G$ form a mate for two pairs
  of adjunctions $F \dashv G$ and $F' \dashv G'$, if the following diagram commutes:
  
  \begin{center}
    \begin{tikzcd}
      {Hom(F'X, Y)} \arrow[rr, dotted, "\simeq"] \arrow[d, "{Hom(\alpha_X, Y)}"'] &  & {Hom(X, G'Y)} \arrow[d, "{Hom(X, \beta_Y)}"] \\
      {Hom(FX, Y)} \arrow[rr, dotted, "\simeq"]                                   &  & {Hom(X, GY)}                                
    \end{tikzcd}
  \end{center}
\end{definition}

This definition is not very convenient because it is defined
via hom-set(oid)s. The situation described in \Cref{sec:ulevel,sec:adjoint} recurs,
and the two natural isomorphisms need to be composed by lifting functors in order to
be well-typed.  As before, there is another definition which does not depend on
hom-sets.
\begin{definition}\label{def:mate:2}
  For functors $F, F' : \mathcal{C} \Rightarrow \mathcal{D}$ and
  $G, G' : \mathcal{D} \Rightarrow \mathcal{C}$, two natural transformation
  $\alpha : F \Rightarrow F'$ and $\beta : G' \Rightarrow G$ form a mate for two pairs
  of adjunctions $(\eta, \epsilon) : F \dashv G$ and
  $(\eta', \epsilon') : F' \dashv G'$, if the following two diagrams commute:
  \begin{center}
    \begin{tikzcd}
      1_{\mathcal{C}} \arrow[r, "\eta"] \arrow[d, "\eta'"'] & GF \arrow[d, "G\alpha"] & FG' \arrow[r, "\alpha G'"] \arrow[d, "F\beta"'] & F'G' \arrow[d, "\epsilon'"] \\
      G'F' \arrow[r, "\beta F'"]                        & GF'                     & FG \arrow[r, "\epsilon"]                        & 1_{\mathcal{D}}                
    \end{tikzcd}
  \end{center}
\end{definition}

Both definitions are equivalent~\cite{maclane:71}, but \Cref{def:mate:2} is simpler
to work with in our setting.

From here, it is straightforward to see that our definition of closed monoidal 
category is equivalent
to the previous ones. We need to show \Cref{def:cmc:4} is equivalent to
requiring $Hom(Y \otimes X, Z) \simeq Hom(Y, [X, Z])$ to be natural in $X$, $Y$,
and $Z$. Since we require $- \otimes X \dashv [X ,-]$ for any object $X$, this
requirement is equivalent to naturality of $Y$ and $Z$. Moreover, the
naturality of $X$ is ensured by the mate condition, due to \Cref{def:mate:1}.

\subsection{Morphism Equality over Natural Isomorphism}\label{sec:lowlevel}

Our experience with monoidal and closed monoidal categories can be generalized into a guideline. We
find that in general, characterization in morphism equalities (such as triangle
identities in \Cref{def:adj:2}) is better than one in natural isomorphisms (such as the
natural isomorphism between hom-sets in \Cref{def:adj:1} and the associativity natural
isomorphism in \Cref{def:monoidal}). The latter can be proved as
a theorem.

We observe that natural isomorphisms tend to be more difficult to type-check, for a
variety of reasons. Similar phenomena are also observed in concepts with higher
structures, e.g. \texttt{Bicategory}, which we encoded directly using morphism
equality to ease the type checking process. 

\subsection{Finite Categories}

Category theorists have developed terminology to talk about the cardinalities (sizes)
of components of a category. In \Cref{sec:ulevel}, we use universe levels to
make size issues explicit. For small categories, since we know both objects and
morphisms ``fit'' in sets, we can use more set-theoretic language.
Among these, ``finiteness'' is of particular importance, especially in its
guise as enabling \textbf{enumeration} and its relation with topoi.

However when we attempt to define finite categories, a problem arises: MLTT does not
give us primitives to count the elements of a type. For example both
\cite{timany_et_al:LIPIcs:2016:6000} and \cite{YorgeyThesis} implement
finiteness as a predicate requiring an isomorphism between a type and
\texttt{Fin $\mathbb{N}$}.  We could also do this,
but that approach has the drawback of (implicitly) putting a canonical order
on elements, which is undesirable\footnote{Propositional truncation could be
used, if we had it, to get around this problem.} It also forces a notion of
equivalence on objects, which does not always exist for any \texttt{Set}.
We do not want finiteness to force us into strictness. We instead base our
definition on adjoint equivalence:
\begin{definition}
  Two categories $\mathcal{C}$ and $\mathcal{D}$ are adjoint equivalent if
  there are two functors $F : \mathcal{C} \to \mathcal{D}$ and $G : \mathcal{D} \to
  \mathcal{C}$ so that they form a pair of adjoint functors $F \dashv G$ and their
  unit and counit natural transformations are isomorphisms. 
\end{definition}
Then a finite category can be defined as follows:
\begin{definition}\label{def:fin-cat}
  A category $\mathcal{C}$ is finite, if it is adjoint equivalent to a finite
  diagram. 
\end{definition}
We could potentially use other notions of equivalence between categories, e.g. strong
equivalence, but adjoint equivalence is special in its smooth interaction with
(co)limits, as will be shown in \Cref{thm:lim-resp-adjeq}. A strong equivalence only
achieves this via its induced adjoint equivalence, so we chose to formulate it more
directly. 

We define a finite diagram using a type family \texttt{Fin : $\mathbb{N}$ → Set}
representing the discrete finite set of natural numbers $\left[0,n-1\right]$
defined in the standard library:
\begin{definition}\label{def:fin-shape}
  Given $n : \mathbb{N}$ as the number of objects and a function
  $|a , b| : \mathbb{N}$ for $a, b : \texttt{Fin x}$, a finite diagram is a category
  with
  \begin{enumerate}
  \item \texttt{Fin n} as objects, and
  \item \texttt{Fin $|a, b|$} as morphisms for \texttt{a, b : Fin n}.
  \end{enumerate}
  if the morphisms satisfy the categorical laws of composition with propositional equality. 
\end{definition}

Intuitively, $|a , b|$ defines an enumeration of the morphisms.  In this category, we
make objects and morphisms discrete, so that propositional equality can be
properly used.

For example, as adjoint equivalence respects equivalence, a contractible
groupoid is always finite. Note that this method could sometimes be challenging:
coming up with such an adjoint equivalence can be difficult and,
in some cases, may require the Axiom of Choice.

Nevertheless, the above definition lets use prove:
\begin{theorem}
  A category with all finite products and equalizers has all finite limits.
\end{theorem}
The proof is constructive, i.e. an algorithm that builds a finite limit from products
and equalizers given any finite diagram. In this theorem, finite limits are described
by functors mapping out of special categories defined in \Cref{def:fin-shape} instead
of the more general \Cref{def:fin-cat}. This theorem at least ensures the sufficiency
of \Cref{def:fin-shape}.

We can then move on to verifying that a finite category as per \Cref{def:fin-cat} can
serve as an index category for a finite limit in the general case. This can be seen
from the following theorem:
\begin{theorem}\label{thm:lim-resp-adjeq}
  Limits respect adjoint equivalence, i.e. if $\mathcal{J}$ is adjoint equivalent to
  $\mathcal{J}'$ with $F : \mathcal{J'} \to \mathcal{J}$, then for a functor $L :
  \mathcal{J} \to \mathcal{C}$, 
  $\underleftarrow{\lim}L = \underleftarrow{\lim}(L \circ F)$. 
\end{theorem}
Combining the two theorems above, we can conclude that \Cref{def:fin-cat} is an adequate
definition of finite categories.  That \Cref{def:fin-cat}  does not involve any
explicit isomorphism between objects and some finite natural numbers is a strength. How much the choice of adjoint equivalence reveals about the inner structure
of a category still remains to be investigated.

\subsection{Local Cartesian Closure of Setoids}

Finally we discuss a complication in proving
that the category of \texttt{Setoids} is locally cartesian closed. This is an
especially interesting theorem to us because base change functors in locally cartesian
categories are left adjoint to the dependent product functors. That implies that
\texttt{Setoids} are a model for dependently typed language.  This theorem
shows some typical extra considerations when proof-relevance and setoids are
involved, and how much implicit equational reasoning we use in classical
settings.

\begin{definition}
  Given a category $\catC$ and its object $X$, a slice category $\catC / X$ has
  \begin{enumerate}
  \item $(Y, f)$ as objects for object $Y$ of $\catC$ and morphism $f : \mor Y X$, 
  \item as a morphism $h : \mor Y Z$ between $(Y, f)$ and $(Z, g)$, so that $g \circ h
    = f$.
  \end{enumerate}
  $X$ is the \emph{base} of $\catC / X$. Given an object $(Y, f)$ in the slice
  category, we often simply refer to it as $f$ as $Y$ can be inferred. 
\end{definition}

\begin{definition}
  A category $\catC$ is cartesian closed when it is closed monoidal with cartesian
  products $\times$ as $\otimes$ and a terminal object as unit. The inner hom $[X,Y]$
  between objects $X$ and $Y$ is the exponential, which is denoted as $Y^X$. 
\end{definition}

\begin{definition}
  A locally cartesian closed category is a category in which all its slice categories
  are cartesian closed. 
\end{definition}

\subsubsection{Classical construction}
Classically, products in the slice category $\catset / A$
are pullbacks in $\catset$. Exponentials can be observed from
the following diagram:
\begin{center}
\begin{tikzcd}
                                                                         &                  & D \arrow[ldd, "h"] \\
B \times_A C \arrow[d, "\pi_1"] \arrow[r, "\pi_2"] \arrow[rru, "\alpha"] & C \arrow[d, "g"] &                    \\
B \arrow[r, "f"]                                                         & A                &                   
\end{tikzcd}
\end{center}
where $B \times_A C$ is an object in $\catset$ and is a pullback of $f$ and $g$. 
From the pullback diagram we want to get an idea of the exponential of $h$ and
$g$, $h^g$.  From the diagram and that $\alpha$ is a slice morphism, we know
\begin{align*}
  Hom(B \times_A C, D) =& \{ \alpha : B \times_A C \to D \;| \\
  & \forall (b, c) \in B \times_A
  C. h(\alpha(b, c)) = f(b) = g(c) \}
\end{align*}
If $\catset / A$ is cartesian closed, then we can find the exponentials via their right
adjointness to pullbacks. 
Assuming the exponential object $h^g$ is a morphism from $X$ to $A$, adjointness 
insures that the
isomorphism $Hom(B \times_A C, D) \simeq Hom(B, X)$ exists. If we were not
working with a slice category, the left-to-right effect is simple, namely just currying,
\begin{align*}
  b : B \mapsto c: C \mapsto \alpha(b, c)
\end{align*}
However, in the slice category, we must ensure that the coherence condition holds,
i.e. $h(\alpha(b, c)) = f(b) = g(c) \in A$. Thus the exponential in the slice
category must carry $f(b)$ and a function,  so we have
\begin{align*}
  X = \Sigma_{a : A} (g^{-1}(a) \to h^{-1}(a))
\end{align*}
That is, as a set, $X$ is a (dependent) pair where the second component
is a function from the inverse image
of $g$ of $a$ to one of $h$ of $a$.  $Hom(B, X)$ is obtained from $\alpha \in Hom(B
\times_A C, D)$ by:
\begin{align*}
  b: B \mapsto (f(b) : A, c : g^{-1}(f(b)) \mapsto \alpha(b, c))
\end{align*}
The presentation contains many hidden details: we can apply $\alpha$ to $c$ because
$g^{-1}(f(b))$ is a subset of $C$, and we know $\alpha(b, c)$ is in $h^{-1}(f(b))$ because
$h(\alpha(b, c)) = f(b)$. Coherence conditions are elided as they can be recovered from the structure of sets.

\subsubsection{In Setoids}
We cannot directly use this kind of reasoning in
\texttt{Setoids}, as we handle setoid morphisms instead. Thus we need a notion of an inverse image \emph{setoid} which
respects setoid equivalence in the codomain. So for some setoid morphism $f$ with
codomain $A$,  if we
have $a \approx a' : A$, then setoids $f^{-1}(a)$ and $f^{-1}(a')$ should have the same
extensional behaviours. This observation is captured by the following theorem:
\begin{minted}{agda}
inverseImage-transport :
  ∀ {a a′} {f : X ⟶ A} → a A.≈ a′ →
    InverseImage a f → InverseImage a′ f
\end{minted}
where \texttt{f : X ⟶ A} specifies that \texttt{f} is a setoid morphism from setoid
\texttt{X} to setoid \texttt{A}.  \texttt{InverseImage a f} formalizes $f^{-1}(a)$ by
requiring some element \agda{x} of setoid \texttt{X} to satisfy \agda{f x A.≈ a} in
setoid \texttt{A}. Moreover, to formalize $g^{-1}(a) \to h^{-1}(a)$ in
\texttt{Setoids}, it is not enough to just provide a function\linebreak
\agda{i : InverseImage a g → InverseImage a h}, because \linebreak
\texttt{InverseImage} contains a proof of \agda{f x A.≈ a} for some \texttt{x}. We
need an extra coherence condition stating that this proof is irrelevant from
\agda{i}'s perspective. That is, given two \texttt{InverseImage}s with \texttt{x} and
\texttt{y} as the underlying elements of \texttt{X}, if \agda{x ≈ y}, then \agda{i x ≈ i y}.
These two pieces of information are bundled in
\agda{InverseImageMap a g h}, which we use to represent the map between inverse image
setoids $g^{-1}(a) \to
h^{-1}(a)$. Finally we need  the following theorem to
ensure that a \texttt{InverseImageMap} respects $A$'s equivalence:
\begin{minted}{agda}
inverseImageMap-transport : ∀ {a a′}
  {g : C ⟶ A} {h : D ⟶ A} → a A.≈ a′ →
  InverseImageMap a g h →
  InverseImageMap a′ g h
\end{minted}
These definitions and theorems fill in the elided coherence conditions in the
classical settings. We can proceed to define an exponential of 
\texttt{h} and \texttt{g} in \texttt{Setoids / A} as a $\Sigma$ type:
\begin{minted}{agda}
Σ (a : A) (InverseImageMap a g h)
\end{minted}
This type does form a setoid with the corresponding setoid equivalence between \texttt{a} and
the underlying map of \texttt{InverseImageMap}, which is the exponential of
\texttt{Setoids / A}. By letting the identity morphism as the terminal object and
pullbacks as products, we can conclude that \texttt{Setoids} is locally cartesian
closed.

%% file: survey.tex
\begin{table*}[]
\caption{Tools and key characteristics of various libraries}\label{tab:tools}

\begin{center}
\begin{tabular}{|l|l|l|l|l|l|}
\hline
libraries                                                & proof assistants         & foundation             & hom-setoids  & proof-relevant & LoC$\dagger$ \\ \hline
Ours                                                     & Agda 2.6.1               & MLTT                   & $\checkmark$ & $\checkmark$   & 23998        \\
\cite{copumpkin}                                         & Agda 2.5.2               & MLTT + K + irrelevance & $\checkmark$ & \xmark         & 11770        \\
\cite{timany_et_al:LIPIcs:2016:6000}                     & Coq 8.11.1               & CIC                    & \xmark       & \xmark         & 14711        \\
\cite{coq:category}                                      & Coq 8.10.2               & CIC                    & $\checkmark$ & $\checkmark$   & 23003        \\
\cite{huet2000constructive}                              & Coq 8.12.0               & CIC                    & $\checkmark$ & $\checkmark$   & 7879         \\
\cite{UniMath,ahrens_kapulkin_shulman_2015}              & Coq 8.12.0               & HoTT                   & \xmark       & $\checkmark$   & 96366        \\
\cite{10.1007/978-3-319-08970-6_18}                      & Hoq 8.12$\dagger\dagger$ & HoTT with HIT          & \xmark       & $\checkmark$   & 10604        \\
\cite{The_mathlib_Community_2020}                        & Lean                     & CIC                    & \xmark       & \xmark         & 14975        \\
\cite{Category3-AFP,MonoidalCategory-AFP,Bicategory-AFP} & Isabelle                 & HOL                    & \xmark       & \xmark         & 82782        \\\hline
\end{tabular}

\end{center}

\begin{itemize}
\item[$\dagger$] The lines of code are counted by \texttt{cloc} of Al Danial and code in
Isabelle is counted by \texttt{wc}, because \texttt{cloc} does not recognize
Isabelle. The lines of code might include documentation text. Only folders directly
related to category theory are counted.
\item[$\dagger\dagger$] Hoq is a a modified version of Coq which implements a part of
  HoTT.
\end{itemize}

\end{table*}

\begin{table*}[]
\caption{Feature comparison (part 1)}\label{tab:comp:1}

\begin{center}
\begin{tabular}{|>{\hangindent=2em}p{7cm}|l|l|l|l|l|l|l|l|l|}
\hline
Features                                 & Ours        & \cite{copumpkin}    & \cite{timany_et_al:LIPIcs:2016:6000}  & \cite{coq:category}    & \cite{huet2000constructive}       & \cite{UniMath,ahrens_kapulkin_shulman_2015}      & \cite{10.1007/978-3-319-08970-6_18}          & \cite{The_mathlib_Community_2020}  & \cite{Category3-AFP,MonoidalCategory-AFP,Bicategory-AFP}    \\ \hline
basic structures:                                &              &              &              &              &              &              &              &              & \\
initial / terminal                               & $\checkmark$ & $\checkmark$ & $\checkmark$ & $\checkmark$ & $\checkmark$ & $\checkmark$ & $\checkmark$ & $\checkmark$ & $\checkmark$ \\
product / coproduct                              & $\checkmark$ & $\checkmark$ & $\checkmark$ &              & $\checkmark$ & $\checkmark$ &              & $\checkmark$ & $\checkmark$ \\
limit / colimit                                  & $\checkmark$ & $\checkmark$ & $\checkmark$ & $\checkmark$ & $\checkmark$ & $\checkmark$ & $\checkmark$ & $\checkmark$ & $\checkmark$ \\
end / coend                                      & $\checkmark$ & $\checkmark$ &              & $\checkmark$ &              & $\checkmark$ &              &              & \\
exponential                                      & $\checkmark$ & $\checkmark$ & $\checkmark$ &              & $\checkmark$ & $\checkmark$ &              &              & \\ \hline
categorical structures:                          &              &              &              &              &              &              &              &              & \\
product / coproduct$\dagger$                     & $\checkmark$ & $\checkmark$ & $\checkmark$ & $\checkmark$ & \hcheckmark  & \hcheckmark  & \hcheckmark  & $\checkmark$ & \hcheckmark  \\
comma category                                   & $\checkmark$ & $\checkmark$ & $\checkmark$ & $\checkmark$ & $\checkmark$ & $\checkmark$ & $\checkmark$ & $\checkmark$ & \\
cartesian category                               & $\checkmark$ & $\checkmark$ & $\checkmark$ & $\checkmark$ & $\checkmark$ &              &              & $\checkmark$ & \\
closed category                                  & $\checkmark$ &              &              &              &              &              &              &              & \\
CCC                                              & $\checkmark$ & $\checkmark$ & $\checkmark$ & $\checkmark$ & $\checkmark$ &              &              & $\checkmark$ & \\
LCCC                                             & $\checkmark$ &              & $\checkmark$ &              &              &              &              &              & \\
biCCC                                            & $\checkmark$ &              &              & $\checkmark$ &              &              &              &              & \\
rig category                                     & $\checkmark$ & $\checkmark$ &              &              &              & $\checkmark$ &              &              & \\
topos                                            & $\checkmark$ &              & $\checkmark$ &              &              &              &              &              & \\
Grothendieck topos                               &              &              &              &              &              & $\checkmark$ &              &              & \\
Eilenberg Moore                                  & $\checkmark$ &              &              & $\checkmark$ &              &              &              & $\checkmark$ & \\
Kleisli                                          & $\checkmark$ &              &              & $\checkmark$ &              & $\checkmark$ &              & $\checkmark$ & \\ \hline
monoidal category                                & $\checkmark$ &              &              & $\checkmark$ &              & $\checkmark$ &              & $\checkmark$ & $\checkmark$ \\
Kelly's coherence~\cite{KELLY1964397}            & $\checkmark$ &              &              &              &              &              &              &              & $\checkmark$ \\
closed monoidal category                         & $\checkmark$ &              &              &              &              &              &              & $\checkmark$ & \\
closed monoidal categories are closed categories & $\checkmark$ &              &              &              &              &              &              &              & \\
braided monoidal category                        & $\checkmark$ & $\checkmark$ &              & $\checkmark$ &              & $\checkmark$ &              & $\checkmark$ & \\
symmetric monoidal category                      & $\checkmark$ & $\checkmark$ &              & $\checkmark$ &              &              &              & $\checkmark$ & \\
traced monoidal category                         & $\checkmark$ & $\checkmark$ &              &              &              &              &              &              & \\
lax monoidal functor                             & $\checkmark$ &              &              & $\checkmark$ &              & $\checkmark$ &              & $\checkmark$ & $\checkmark$\\
strong monoidal functor                          & $\checkmark$ &              &              & $\checkmark$ &              & $\checkmark$ &              &              & \\ \hline
instances:                                       &              &              &              &              &              &              &              &              & \\
\textbf{Cats}                                    & $\checkmark$ & $\checkmark$ & $\checkmark$ & $\checkmark$ & $\checkmark$ & $\checkmark$ & $\checkmark$ & $\checkmark$ & \\
\textbf{Set(oid)s}                               & $\checkmark$ & $\checkmark$ & $\checkmark$ & $\checkmark$ & $\checkmark$ & $\checkmark$ & $\checkmark$ & $\checkmark$ & $\checkmark$ \\
\textbf{Setoids} are complete / cocomplete       & $\checkmark$ &              & $\checkmark$ & $\checkmark$ & $\checkmark$ &              &              &              & \\
\textbf{Setoids} are cartesian closed            & $\checkmark$ &              & $\checkmark$ & $\checkmark$ & $\checkmark$ &              &              &              & \\
\textbf{Setoids} are locally cartesian closed    & $\checkmark$ &              & $\checkmark$ &              &              &              &              &              & \\
simplicial set                                   & $\checkmark$ &              &              &              &              & $\checkmark$ &              &              & \\ \hline
functor                                          & $\checkmark$ & $\checkmark$ & $\checkmark$ & $\checkmark$ & $\checkmark$ & $\checkmark$ & $\checkmark$ & $\checkmark$ & $\checkmark$ \\
(co)limit functor                                & $\checkmark$ &              & $\checkmark$ &              &              & $\checkmark$ & $\checkmark$ &              & $\checkmark$ \\
Hom functor                                      & $\checkmark$ & $\checkmark$ & $\checkmark$ & $\checkmark$ & $\checkmark$ & $\checkmark$ & $\checkmark$ & $\checkmark$ & $\checkmark$ \\
Hom functors preserve limits                     & $\checkmark$ &              & $\checkmark$ &              & $\checkmark$ &              &              &              & \\
T-algebra                                        & $\checkmark$ & $\checkmark$ & $\checkmark$ & $\checkmark$ &              & $\checkmark$ &              &              & \\
Lambek's lemma                                   & $\checkmark$ & $\checkmark$ &              &              &              & $\checkmark$ &              &              & \\ \hline
natural transformation                           & $\checkmark$ & $\checkmark$ & $\checkmark$ & $\checkmark$ & $\checkmark$ & $\checkmark$ & $\checkmark$ & $\checkmark$ & $\checkmark$ \\
dinatural transformation                         & $\checkmark$ & $\checkmark$ &              & $\checkmark$ &              &              &              &              & \\ \hline
enriched category                                & $\checkmark$ & $\checkmark$ &              & $\checkmark$ &              & $\checkmark$ &              &              & \\
2-category                                       & $\checkmark$ & $\checkmark$ &              &              &              & $\checkmark$ &              &              & $\checkmark$ \\
bicategory                                       & $\checkmark$ & $\checkmark$ &              & $\checkmark$ &              & $\checkmark$ &              &              & $\checkmark$ \\
pseudofunctor                                    & $\checkmark$ & $\checkmark$ &              &              &              & $\checkmark$ & $\checkmark$ &              & $\checkmark$ \\ \hline
Yoneda lemma                                     & $\checkmark$ & $\checkmark$ & $\checkmark$ & $\checkmark$ & $\checkmark$ & $\checkmark$ & $\checkmark$ & $\checkmark$ & $\checkmark$ \\ \hline
\end{tabular}
\end{center}

\begin{enumerate}
\item[$\dagger$] \hcheckmark indicates that these libraries only implement product
categories.
\end{enumerate}
\end{table*}

\begin{table*}[bt!]
\caption{Feature comparison (part 2)}\label{tab:comp:2}

\begin{center}
\begin{tabular}{|>{\hangindent=2em}p{7cm}|l|l|l|l|l|l|l|l|l|}
\hline
Features                                 & Ours        & \cite{copumpkin}    & \cite{timany_et_al:LIPIcs:2016:6000}  & \cite{coq:category}    & \cite{huet2000constructive}       & \cite{UniMath,ahrens_kapulkin_shulman_2015}      & \cite{10.1007/978-3-319-08970-6_18}          & \cite{The_mathlib_Community_2020}  & \cite{Category3-AFP,MonoidalCategory-AFP,Bicategory-AFP}    \\ \hline
Grothendieck construction                        & $\checkmark$ & $\checkmark$ &              &              &              &              & $\checkmark$ & \hcheckmark$\dagger$$\dagger$ &              \\ \hline
presheaves                                       & $\checkmark$ & $\checkmark$ & $\checkmark$ & $\checkmark$ &              & $\checkmark$ &              &  $\checkmark$                  &              \\
are complete / cocomplete                        & $\checkmark$ &              & $\checkmark$ &              &              &              &              &  $\checkmark$                  &              \\
are cartesian closed                             & $\checkmark$ &              & $\checkmark$ &              &              &              &              &                               &              \\
are topos                                        &              &              & $\checkmark$ &              &              &              &              &                               &              \\ \hline
adjoint functors                                 & $\checkmark$ & $\checkmark$ & $\checkmark$ & $\checkmark$ & $\checkmark$ & $\checkmark$ & $\checkmark$ & $\checkmark$                  & $\checkmark$ \\
adjoint composition                              & $\checkmark$ & $\checkmark$ & $\checkmark$ & $\checkmark$ &              &              & $\checkmark$ & $\checkmark$                  & $\checkmark$ \\
Right(left) adjoints preserve (co)limits         & $\checkmark$ &              & $\checkmark$ &              & $\checkmark$ & $\checkmark$ &              & $\checkmark$                  & $\checkmark$ \\
Adjoint functors induce monads                   & $\checkmark$ &              &              & $\checkmark$ &              & $\checkmark$ &              & $\checkmark$                  &              \\
(Co)limit functors are left(right)
  adjoint to diagonal functor$\dagger$           & $\checkmark$ &              & $\checkmark$ & \hcheckmark  &              & \hcheckmark  & $\checkmark$ &                               & $\checkmark$ \\
mate (conjugate)                                 & $\checkmark$ &              &              &              &              & $\checkmark$ &              &                               &              \\
adjoint functor theorem                          & $\checkmark$ &              & $\checkmark$ &              & $\checkmark$ &              &              &                               &              \\ \hline
Kan extension                                    & $\checkmark$ & $\checkmark$ & $\checkmark$ & $\checkmark$ &              & $\checkmark$ & $\checkmark$ &                               &              \\
(Co)limit is kan                                 & $\checkmark$ &              & $\checkmark$ & $\checkmark$ &              & $\checkmark$ &              &                               &              \\
Kan extensions are preserved by adjoint functors &              &              & $\checkmark$ &              &              &              &              &                               &              \\ \hline
Rezk completion                                  &              &              &              &              &              & $\checkmark$ &              &                               &              \\ \hline
\end{tabular}
\end{center}

\begin{enumerate}
\item[$\dagger$]  \hcheckmark indicates that these libraries only show a special case of the
  theorem.
\item[$\dagger$$\dagger$] \hcheckmark indicates that \cite{The_mathlib_Community_2020}
  only implements the category of elements.
\end{enumerate}

\end{table*}

%% file: discussion.tex
\section{Discussion}\label{sec:discuss}

The previous section detailed decisions that lie in the intersection of
category theory and formalization in type theory, here we document
software engineering decisions as well as comment on efficiency issues.

\subsection{Module Structure}

The previous library favoured a flat module structure, we use a deeper hierarchy,
and thus fewer top-level modules. We use the following principles as a guide:
\begin{enumerate}
\item Important concepts have their top level modules. For example, \texttt{Category},
  \texttt{Object}, \texttt{Morphism}, \texttt{Diagram}, \linebreak \texttt{Functor},
  \texttt{NaturalTransformation}, \texttt{Kan}, \texttt{Monad} and
  \texttt{Adjoint} belong to this category.
\item Different flavours of category theory are also on the top level:
  \texttt{Category}, \texttt{Enriched}, \texttt{Bicategory} and \linebreak
  \texttt{Minus2-Category} contain the definitions and properties of categories,
  enriched categories, bicategories and -2-categories,
  respectively. \texttt{Pseudofunctor} contains the instances of pseudofunctors.
\end{enumerate}
Submodules also follow conventions so that definitions and properties are easier to
locate.
\begin{enumerate}
\item \texttt{*.Instance} contains instances of some concept. For example, the
  category of all setoids is defined in \linebreak\texttt{Category.Instance}.  Generally,
  only instances that are re-used in the library itself (making them ``special'')
  are defined.
\item \texttt{*.Construction} contains instances induced from some input. The difference
  with \texttt{*.Instance}  is that \linebreak\texttt{*.Construction} takes parameters
  beyond just \texttt{Level}s. For example, the Kleisli category of a monad is defined
  in \texttt{Category.Construction}.

\item \texttt{*.Properties} contains properties of the corresponding concepts.
  
\item \texttt{*.Duality} contains conversions to dual concepts (see \Cref{sec:duality}).
\end{enumerate}
This module structure was inspired by a recent restructuring of Agda's standard
library along similar lines, which we believe helps users find what they need faster.

\subsection{Hierarchy of Concepts}

Similar to
\cite{10.1007/978-3-319-08970-6_18,DBLP:conf/tphol/GarillotGMR09,DBLP:journals/mscs/SpittersW11,DBLP:journals/jsc/GeuversPWZ02},
we need to decide how concepts are organized. Unlike Coq, which many cited works are
based on, Agda does not have features like canonical structures or hint based
programming.  But, like the standard library, we do not wish to use type classes.
One reason is performance: at this moment, type classes in Agda are
fairly slow (compared to, say, Coq), potentially penalizing downstream librairies
and end users. Nevertheless, we still need to organize our library so that concepts
can be found.

At the lowest level, we rely on records and unification.
There are typically two choices to represent a concept: predicates or
structures. A predicate has the data ``unbundled''; it expresses an ``is-a''
relation. A structure on the other hand is ``bundled'' and expresses a ``has-a''
relation. The previous library, and many other implementations too, chose
to either bundle or unbundle. From a type-theoretic perspective, this
choice is irrelevant, but is nevertheless quite important from a usability
perspective. It is even possible to automatically map from one style to 
another~\cite{unbundling}; unfortunately, such mapping is meta-theoretical in
current Agda. As such a choice is unforced, we decided to implement both.

\paragraph{Wrapping Predicates}

Structures are obtained by wrapping predicates. 
Influenced by the previous library~\cite{copumpkin},
many concepts related to \texttt{Category} are represented as predicates:
\begin{minted}{agda}
record Monoidal {o ℓ e} (C : Category o ℓ e)
    : Set (o ⊔ ℓ ⊔ e) where
\end{minted}
It asserts that \texttt{C} \textit{is} a monoidal category. At other times,
e.g. when working with two monoidal categories,
we want to represent monoidal categories as a structure. We provide definitions
in both styles:
\begin{minted}{agda}
record MonoidalCategory o ℓ e
    : Set (suc (o ⊔ ℓ ⊔ e)) where
  field
    U        : Category o ℓ e
    monoidal : Monoidal U
\end{minted}
\texttt{U} stands for ``underlying''. This allows us to define (lax) monoidal functors,
which are functors preserving the \linebreak monoidal structure:
\begin{minted}{agda}
record MonoidalFunctor
  (C : MonoidalCategory o ℓ e)
  (D : MonoidalCategory o′ ℓ′ e′)
  : Set (o ⊔ ℓ ⊔ e ⊔ o′ ⊔ ℓ′ ⊔ e′) where
\end{minted}
The alternative formulation using the predicate representation is more
verbose:
\begin{minted}{agda}
record MonoidalFunctor′
  {C : Category o ℓ e} {D : Category o′ ℓ′ e′}
  (MC : Monoidal C)    (MD : Monoidal D)
  : Set (o ⊔ ℓ ⊔ e ⊔ o′ ⊔ ℓ′ ⊔ e′) where
\end{minted}
When working with monoidal functors, we do not mean to assert that some
category is monoidal but rather want to refer to some structured
category as a whole.

In general, definitions in the structure style
are defined in modules of the form \texttt{*.Structure}. As the previous
library used the predicate style, we started our in that style as
well and then provided wrapped structure versions. As a rule of thumb, when
working with one particular concept, we often use the predicate style so that the
conclusions can be easily accessed by both styles. For example, we formulate theorems about
monoidal categories using the predicate style.

The paper~\cite{unbundling} further discusses (un)bundling of definitions, along with
tools for moving between the two equivalent styles.

\paragraph{Choosing Predicates}

Next we use cartesian products to illustrate how we design predicate
formulations.  We have the following structure-kind definition for products:
\begin{minted}{agda}
record Product (A B : Obj) : Set _ where
\end{minted}
The record contains projections, product morphisms, and necessary laws for a
product. This definition works very well when we work on \emph{one} category. However,
when we work with two categories, then we need a predicate version:
\begin{minted}{agda}
record IsProduct {A B P}
  (π₁ : P ⇒ A) (π₂ : P ⇒ B) : Set _ where
\end{minted}
In the arguments, \agda{P} represents the product of \agda{A} and \agda{B}, and
\agda{π₁} and \agda{π₂} are the projections. It is possible to have a slightly
different predicate definition:
\begin{minted}{agda}
record IsProduct′ {A B P}
  (π₁ : P ⇒ A) (π₂ : P ⇒ B)
  (⟨_,_⟩ : C ⇒ A → C ⇒ B → C ⇒ P) :
  Set _ where
\end{minted}
where \agda{⟨ f , g ⟩} denotes the product of morphism \agda{f} and \agda{g}. We did
not choose this form because \agda{⟨ f , g ⟩} is uniquely determined by \agda{π₁} and
\agda{π₂}! That is, even if \agda{IsProduct} allows a ``different'' \agda{⟨ f , g ⟩′},
they are provably equivalent. In general, when formulating concepts
defined by universal properties, we can omit the universal part in the predicate form
due to uniqueness. 

\subsection{Efficiency}

Basic category theory typechecks very quickly, both online (via Emacs) and
offline (via calling the \texttt{agda} compiler). But for ``deeper''
category theory, such as properties associated to the Yoneda lemma and
properties of Bicategories, typechecking gets noticeably slower and
memory use goes up.  One of the culprits is the module style as documented
in Section~\ref{sec:encap}: such modules are copied and rechecked, which
is quite inefficient. This is why when we use local modules (either
\texttt{private} or in \texttt{where} clauses) we qualify them with
\texttt{using} to only copy the parts we need.

Unfortunately that same trick does not work for global \texttt{open import}
(for sound reasons). Agda's \texttt{.agdai} file format is very information-rich
(i.e. the files are quite large), and full transitive dependencies must be
read. Splitting developments into smaller files to minimize the dependency
tree has lead to substantial improvements in the compilation time and
memory use of the full library. The downside is that some usability
features have had to be sequestered into sub-modules that are then imported
on an as-needed basis.

%% file: related.tex
\section{Related Work}\label{sec:related}

\Cref{tab:tools} gives a list of formalized libraries of category theory.  For
each we specify the proof assistant, the foundation, lines of code and whether
it uses hom-setoids and is proof-relevant. In~\Cref{tab:comp:1,tab:comp:2}, we
compare a list of features implement by these libraries. 

We have ported all definitions and theorems from~\cite{copumpkin}, 
except those requiring UIP or axiom K. We reuse \cite{copumpkin} as much as we can.  We also extend it with many new
definitions and new theorems, as shown in
\Cref{tab:comp:1,tab:comp:2} (more than twice as much material).
Moreover, since we turn on the \opt{safe}
flag, we do not have \texttt{postulate}s in our code base. This helps us to avoid
inheriting a postulated unsound axiom~\cite{lmcs:1045},
which would, for example, let us incorrectly mix relevance and irrelevance,
including ``recovering'' a relevant value from an irrelevant one.

From~\Cref{tab:tools}, we can see that much effort has been spent
in Coq (or its Hoq dialect) on category theory.  The reason for the multiple efforts
can be seen when comparing the versions, and foundations used. These libraries also
vary in their design and organization.   Some believe that
Coq's tactics and hint databases provide a significant boost in the
productivity of formalizations. We suspect that this may be somewhat illusory, as
the explicit equational proofs in $n$-category theory (which can be automated via
tactics) tend to turn up as data in $n+1$-category theory, and then no longer
avoidable.  \cite{coq:category} stands out by its use of other Coq mechanisms,
such as type classes, rather than record or $\Sigma$ types, for structuring of
the development.

Like us, \cite{coq:category,huet2000constructive} use hom-setoids and
proof-relevance. Unfortunately, \cite{coq:category} has not been described
in a paper, so we do not know what lessons the authors learned from their
experience. \cite{huet2000constructive} was a smaller scale but pioneering
effort that taught us the basics of formalizing category theory in MLTT, but
not the kinds of design decisions we faced here.

Compared to other developments in Coq, \cite{10.1007/978-3-319-08970-6_18,UniMath} are
special: they build category theory in HoTT.  \cite{UniMath} focuses
more on fundamental constructions. It does not use any feature beyond the primitive
type constructors like $\Sigma$ and $\Pi$. By contrast,
\cite{10.1007/978-3-319-08970-6_18} experiments with the 
use of various HoTT ideas, and therefore is more
permissive. It uses extended features like records and higher inductive types (HITs).
Working in HoTT has some advantages.  First, if one understands hom-sets to be
literally classical sets, rendered as \texttt{hSet}s in HoTT, this is straightforward.
In HoTT this also implies that \texttt{hSet}s have unique identity proofs, which
make their equational proofs proof-irrelevant, which is closer to
the set-based understanding of classical category theory.
Second, HoTT has a very natural way of expressing universal
properties. Using Martin-L\"of type theories, e.g. ours,
\cite{copumpkin,coq:category,huet2000constructive,timany_et_al:LIPIcs:2016:6000},
universal properties are usually stated in two parts: a universal part returning a
morphism and a uniqueness part equating morphisms from the universal part. In HoTT,
this can be expressed compactly as constructing a contractible morphism. Third, since
HoTT supports the univalence, one can conflate isomorphisms and equalities. In
both libraries, categories are defined with an additional law stating that
isomorphic objects are equal, which provides a way to handle equal objects in a
category which ours does not have. 

The mathematical library of Lean~\cite{de2015lean},
mathlib~\cite{The_mathlib_Community_2020}, also implements some category
theory\footnote{This library is being actively developed. Our survey is
valid as of mid-September 2020 and does not consider the open PRs to
the main library.}. As Lean has \linebreak proof-irrelevance built in and mathlib uses propositional
equality, its category theory library is very classical.


Category theory has also been formalized in
Nuprl~\cite{10.5555/10510}, Idris~\cite{brady2013idris} and Isabelle~\cite{10.5555/1791547}. 
Due to space limitation,
we are not able to fully survey all of them. We refer interested readers to
\cite{10.1007/978-3-319-08970-6_18} and the Coq discourse forum\footnote{ 
  \url{https://coq.discourse.group/t/survey-of-category-theory-in-coq/371/4}.}
for a more thorough list of formalizations of category theory.